\documentclass[aps,prd,twocolumn,showpacs,superscriptaddress,nofootinbib,10pt]{revtex4-2}

\usepackage{acronym}
\usepackage{amsfonts}
\usepackage{amsmath}
\usepackage{amssymb}
\usepackage{bm}
\usepackage{cases}
\usepackage{color}
\usepackage{comment}
\usepackage[dvipsnames]{xcolor}
\usepackage{enumerate}
\usepackage{epsfig}
\usepackage{etoolbox}
\usepackage{fancyref}
\usepackage{fancyhdr}
\usepackage[T1]{fontenc} 
\usepackage{graphicx}
\usepackage{hyperref}
\usepackage{indentfirst}
\usepackage{latexsym}
\usepackage{mathrsfs}
\usepackage{mciteplus}
\usepackage{multirow}
\usepackage{rotating}
\usepackage{scrextend}
\usepackage{soul}
\usepackage{subfigure}
\usepackage{verbatim}

\newcommand{\jeven}{\mbox{\rm\j}}

\begin{document}

\title[Master Functions and Equations for Perturbations of Vacuum Spherically-Symmetric Spacetimes]{Master Functions and Equations for Perturbations of Vacuum Spherically-Symmetric Spacetimes}

\author{Michele Lenzi}
\affiliation{Institut de Ci\`encies de l'Espai (ICE, CSIC), Campus UAB, Carrer de Can Magrans s/n, 08193 Cerdanyola del Vall\`es, Spain}
\affiliation{Institut d'Estudis Espacials de Catalunya (IEEC), Edifici Nexus, Carrer del Gran Capit\`a 2-4, despatx 201, 08034 Barcelona, Spain}
\affiliation{Dipartimento di Fisica e Astronomia, Universit\`a di Bologna, via Irnerio 46, 40126 Bologna, Italy}

\author{Carlos F. Sopuerta}
\affiliation{Institut de Ci\`encies de l'Espai (ICE, CSIC), Campus UAB, Carrer de Can Magrans s/n, 08193 Cerdanyola del Vall\`es, Spain}
\affiliation{Institut d'Estudis Espacials de Catalunya (IEEC), Edifici Nexus, Carrer del Gran Capit\`a 2-4, despatx 201, 08034 Barcelona, Spain}

\begin{abstract}
Perturbation theory of vacuum spherically-symmetric spacetimes is a crucial tool to understand the dynamics of black hole perturbations.   
Spherical symmetry allows for an expansion of the perturbations in scalar, vector, and tensor harmonics. The resulting perturbative equations are decoupled for modes with different parity and different harmonic numbers. Moreover, for each harmonic and parity, the equations for the perturbations can be decoupled in terms of (gauge-invariant) master functions that satisfy $1+1$ wave equations. 
By working in a completely general perturbative gauge, in this paper we study what is the most general master function that is linear in the metric perturbations and their first-order derivatives and satisfies a wave equation with a potential. 
The outcome of the study is that for each parity we have two branches of solutions with similar features. One of the branches includes the known results: In the odd-parity case, the most general master function is an arbitrary linear combination of the Regge-Wheeler and the Cunningham-Price-Moncrief master functions whereas in the even-parity case it is an arbitrary linear combination of the Zerilli master function and another master function that is new to our knowledge. The other branch is very different since it includes an infinite collection of potentials which in turn lead to an independent collection master of functions which depend on the potential. The allowed potentials satisfy a non-linear ordinary differential equation. Finally, all the allowed master functions are gauge invariant and can be written in a fully covariant form.
\end{abstract}

\maketitle

\section{Introduction}
Spacetime perturbation theory is one of the main tools in General Relativity to describe physical systems and make reliable predictions about their characteristics and dynamical behavior. It has been applied essentially to all the main problems in the area of relativistic astrophysics and cosmology: From the origin and growth of cosmological structures and the cosmic microwave background~\cite{Bardeen:1980PhRvD..22.1882B,Kodama:1984PThPS..78....1K,Bruni:1989PhRvD..40.1804E,Hu:1994jd} to phenomena involving relativistic stars and black holes~\cite{Nollert:1999re,Kokkotas:1999bd,Andersson:2000mf,Kokkotas:2003mh}; including gravitational wave generation and propagation~\cite{Peters:1964PhRv..136.1224P,Misner:1973cw,Blanchet:2013haa,Tiec:2014lba} (see~\cite{Davis:1971gg,Dreyer:2003bv,Sasaki:2003xr,Ferrari:2007dd,Sathyaprakash:2009xs,Tiec:2014lba,Berti:2018vdi} for applications of black hole perturbation theory to gravitational wave astronomy). There are also many applications of black hole perturbation theory that have implications for fundamental physics (see~\cite{Cardoso:2012qm,Brito:2015oca,Barack:2018yly} for reviews).

The long experience in relativistic perturbation theory tells us that it is a very powerful tool, in part because starting from very simplified situations, encoded in what is known as the {\em background} spacetime, it has shown to provide results and predictions that many times go beyond the range of applicability that one may expect from back-of-the-envelope estimates.
For example, this has happened when we have compared the outcome of full Numerical Relativity simulations with post-Newtonian/Minkowskian theories and relativistic perturbation theory.

In this work we focus on relativistic perturbation theory of vacuum spherically-symmetric spacetimes. We allow for the presence of a cosmological constant which, as we are going to see, does not make our computations much more complicated since it can be incorporated into our equations in a relatively simple way. This means that our study includes the dynamics of perturbations around Schwarzschild, Schwarzschild-de Sitter, and Schwarzschild-anti de Sitter spacetimes, including also the associated maximally-symmetric spacetimes: Minkowski, de Sitter, and Anti-de Sitter respectively. It is well known that  perturbations of spherically symmetric spacetimes can be decomposed in spherical harmonics, in such a way that the associated equations for the different harmonic modes decouple.
Moreover, the equations for different parity harmonics also decouple, i.e. odd-parity (axial) modes can be treated separately from even-parity (polar) modes. 
Another crucial feature of the theory of perturbations of spherically symmetric spacetimes is that, for each harmonic and parity, it is possible to construct {\em master functions}, made out of linear combinations of the metric perturbations and their first-order derivatives, that satisfy $1+1$ wave (master) equations decoupled from the rest of metric perturbations.
The characteristic curves of the wave operator of the master equations are fully determined by the background (Lorentzian) metric.  In some perturbative gauges, it is even possible to reconstruct all the metric perturbations in terms of the solutions of the master equations, something that is required in some problems, like in the self-force program~\cite{Poisson:2011nh,Barack:2009ux,Barack:2018yvs}.

In the case of perturbations of the Schwarzschild metric, the first steps~\cite{Schwarzschild:1916uq} were already taken in the 1950s by Regge and Wheeler~\cite{Regge:1957td} (see also~\cite{Vishveshwara:1970cc,Vishveshwara:1970zz}), who managed to decouple the equations for odd-parity perturbations in the gauge named after them (also the master function and equation are named after them).
However, it took a while until the same was done for even-parity perturbations, when Zerilli found the way to decouple the perturbative equations~\cite{Zerilli:1970se,Zerilli:1970la} (see also~\cite{Moncrief:1974vm}). 
Starting from these pioneering works, there have been many interesting developments in perturbation theory of spherically symmetric spacetimes: gauge-invariant and/or explicitly-covariant formalisms~\cite{Gerlach:1979rw,Gerlach:1980tx,Sarbach:2001qq,Clarkson:2002jz} (see~\cite{Mukohyama:2000ui,Kodama:2003jz} for the case of D-dimensional maximally-symmetric spacetimes); quasinormal modes~\cite{Ferrari:1984zz,Nollert:1999re,Kokkotas:1999bd,Berti:2009kk,Konoplya:2011qq} (for $D$ dimensions see~\cite{Natario:2004jd}); stability of dynamics of the perturbations~\cite{Kay:1987ax,Kay:1988mu}; stability of black holes in de Sitter Space~\cite{Mellor:1989ac}; etc.
For reviews on non-rotating black hole perturbation theory see~\cite{Chandrasekhar:1992bo,Mino:1997bx,Sarbach:2001qq,Nagar:2005ea,Martel:2005ir} (for second-order perturbations see~\cite{Gleiser:1995gx,Gleiser:1998rw,Garat:2000gp}). On the other hand, similar developments took place for perturbations of rotating black holes~\cite{Teukolsky:1972my,Teukolsky:1973ha}.
For studies of the stability of black holes in asymptotically-flat spacetimes see~\cite{Ishibashi:2003ap,Dafermos:2016uzj}.

In this paper, we further investigate the construction of master functions and equations. In particular,  we focus on the following questions: What is the most general master function that decouples the equations for the metric perturbations of spherically-symmetric vacuum spacetimes? And, what are the possible potentials associated with those master functions?  
To answer these questions we assume that the master functions are linear combinations  (with coefficients that depend only on the radial areal coordinate, $r$) of the metric perturbations and their first-order derivatives. Our analysis turns out to be very similar for the odd- and even-parity cases, also leading us to similar conclusions.  
The result we find is that we can distinguish two branches of solutions (for each parity and harmonic mode). The first branch is mostly known: The master functions are linear combinations of two different (linearly independent) master functions. In the odd-parity case they can be taken to be the Regge-Wheeler~\cite{Regge:1957td} and the Cunningham-Price-Moncrief~\cite{Cunningham:1978cp,Cunningham:1979px,Cunningham:1980cp} master functions. In the even-parity case, one of them can be taken to be the well-known Zerilli-Moncrief master function~\cite{Zerilli:1970se,Zerilli:1970la,Moncrief:1974vm}, while the second one, as far as we know, was previously unknown.  
It turns out that, for both parities, the independent master functions can be chosen so that one of them is the time-derivative of the other one.  Regarding the master equations themselves, which in our case are essentially determined by the potential, in the first branch we find the well-known potentials: The Regge-Wheeler potential for odd-parity perturbations and the Zerilli potential for even-parity perturbations. On the other side, the second branch was essentially unknown. To begin with, there are infinite possible potentials, different from the ones already known in the first branch. Actually, the allowed potentials satisfy a non-linear ordinary differential equation. The master functions are again a linear combination (again with coefficients that depend only on $r$) of two independent master functions. In the odd-parity case, they can be taken to be the Cunningham-Price-Moncrief master function and a new one that combines the Cunningham-Price-Moncrief and another gauge-invariant function.  The even-parity case is analogous, the most general master function is a linear combination of the Zerilli-Moncrief master function and another new master function that contains the Zerilli-Moncrief master function and a gauge-invariant variable.

Some remarkable features and consequences coming from this study are: 
(i) All the master functions involved are automatically gauge-invariant. 
(ii) All the master functions and master equations admit a fully covariant form with respect to the 1+1 Lorentzian metric. 
(iii) In the same way that the Regge-Wheeler and Zerilli potentials coincide for the case of a maximally-symmetric background, the equations for the potentials in the second branch also coincide. 
(iv) Our approach constitutes an original and systematic way of searching for master functions and equations without having to resort to look for {\em ad hoc} combinations of the perturbative field equations that yield decoupled master equations.
 
This paper is organized as follows: In Section~\ref{relativistic-perturbation-theory-spherically-symmetric} we introduce all the necessary elements of relativistic perturbation theory for (vacuum) spherically-symmetric spacetimes, including: The elements associated with the background spacetime; the decomposition of the perturbations in spherical harmonics; gauge invariance; and the known master functions in different forms. In Section~\ref{construction-master-functions}, we describe how to obtain the most general master functions and equations satisfying the hypothesis mentioned above. Finally, in Section~\ref{Conclusions-Future-Prospects} we summarize and discuss the results of this paper. We also include two Appendices with key formulae used in this work.
We use geometric units in which $G=c=1$.

\section{Relativistic Perturbation Theory of Spherically-Symmetric Spacetimes} \label{relativistic-perturbation-theory-spherically-symmetric}
In this section we introduce all the ingredients and machinery needed to derive the main results of this work.

\subsection{Basics of Relativistic Perturbation Theory}\label{relativistic-perturbation-theory}

Relativistic Perturbation Theory is usually formulated by assuming the existence of a one-parameter family of spacetimes, $(\mathcal{M}_\lambda, g_\lambda)$, in such a way that the perturbations are constructed as a Taylor expansion of this family around the $\lambda=0$ spacetime (see, e.g.~\cite{Stewart:1974uz,Wald:1984cw}), the {\em background} spacetime, which usually represents an idealized gravitational system, typically with a high degree of symmetry as in our case. Here, we assume the background\footnote{We use a hat to denote quantities associated with the background spacetime, like $\widehat{Q}$.} to be a vacuum (including the cosmological constant) spherically-symmetric spacetime.  Then, the background spacetime metric, $\widehat{g}^{}_{\mu\nu}$, satisfies the vacuum Einstein's field equations including the cosmological constant term:
\begin{eqnarray}
\widehat{G}^{}_{\mu\nu} = \widehat{R}^{}_{\mu\nu} -\frac{1}{2}\widehat{g}^{}_{\mu\nu} \widehat{R} + \Lambda\,\widehat{g}^{}_{\mu\nu} = 0\,,
\label{efes-background}
\end{eqnarray}
where $\widehat{R}^{}_{\mu\nu}$ and $\widehat{G}^{}_{\mu\nu}$ denote the Ricci and Einstein tensors of the background metric respectively, $\widehat{R} = \widehat{g}^{\mu\nu} \widehat{R}^{}_{\mu\nu}$ is the background scalar curvature, and $\Lambda$ is the cosmological constant.

In this framework, the perturbations are defined as the derivative terms of the Taylor series expansion, evaluated on the background. The parameter $\lambda$ controls the strength of the perturbations and in most applications it is a formal parameter without a specific physical meaning, except in some cases in which it is identified with some relevant physical parameter of the system (see, e.g.~\cite{Stewart:1974uz,Bruni:1996im,Sopuerta:2003rg} for more details on the formulation of relativistic perturbation theory).  Since in our case $\lambda$ is a formal parameter, we are going to ignore it from now on for the sake of simplicity.
The physical (perturbed) spacetime, a member of the one-parameter family of spacetimes $(\mathcal{M}_\lambda, g_\lambda)$,  is endowed with a metric $g_{\mu\nu}$ which, once a correspondence between the background spacetime is established, can be constructed to linear order from the background solution $\widehat{g}^{}_{\mu\nu}$ and the  metric perturbations  $h_{\mu\nu}$ ($|h_{\mu\nu}|\ll|\widehat{g}_{\mu\nu}|$) via the relation:
\begin{equation}
g^{}_{\mu\nu} = \widehat{g}^{}_{\mu\nu} + h^{}_{\mu\nu} \,.
\label{fundamental-perturbative-equation}
\end{equation}
For any quantity $Q$, we denote the deviations between the perturbative and background expressions with a $\delta$ in front of the original quantity, i.e. $\delta Q = Q - \widehat{Q}$, where $Q$ is the expression from the perturbed/physical spacetime.  
In this way: $h_{\mu\nu} = \delta g_{\mu\nu} = g_{\mu\nu} - \widehat{g}_{\mu\nu}$. When we expand such a quantity in the different perturbative orders we are actually performing Taylor expansions in the parameter $\lambda$. In this sense, at first-order, the perturbed Christoffel symbols can be written in terms of the metric perturbations $h_{\mu\nu}$ and their covariant derivatives with respect to the background metric (denoted here by a semicolon), as follows:
\begin{eqnarray}
\delta\Gamma_{\mu\nu}^\rho= \frac{1}{2}\widehat{g}^{\rho\sigma}_{}\left( h^{}_{\mu\sigma;\nu} + h^{}_{\nu\sigma;\mu} - h^{}_{\mu\nu;\sigma} \right)\,.
\label{perturbed-Christoffel}
\end{eqnarray}
From the expression of these quantities we deduce that they are tensors with respect to coordinate changes in the background spacetime.  Then, we can write the perturbations of the Riemann tensor in terms of the perturbed Christoffel symbols (which are tensors from the point of view of the background spacetime) as follows
\begin{equation}
\delta R^{\mu}{}^{}_{\nu\rho\sigma} = \delta\Gamma^{\mu}_{\nu\sigma;\rho} -  \delta\Gamma^{\mu}_{\nu\rho;\sigma} = 
2\delta\Gamma^{\mu}_{\nu[\sigma;\rho]} \,.
\label{perturbed-Riemann}
\end{equation}
In the same way, the perturbations of the Ricci tensor can be written in terms of the covariant derivatives of the perturbed Christoffel symbols:
\begin{eqnarray}
\delta R^{}_{\mu\nu} = \delta\Gamma^{\rho}_{\mu\nu;\rho} - \delta\Gamma^{\rho}_{\rho\mu;\nu}\,.
\label{perturbed-Ricci0}
\end{eqnarray}
The Einstein tensor can be decomposed as: $G_{\mu\nu} = \widehat{G}^{}_{\mu\nu} + \delta G_{\mu\nu} = \delta G_{\mu\nu}$, where the second equality holds by virtue of the Einstein field equations satisfied by the background metric [Eq.~\eqref{efes-background}]. The perturbation in the Einstein tensor in terms of the metric perturbations $h_{\mu\nu}$ are
\begin{eqnarray}
\delta G^{}_{\mu\nu} & = & -\frac{1}{2}\widehat{\square}\,\bar{h}^{}_{\mu\nu} -\widehat{R}^{\rho}_{}{}^{}_\mu{}^{\sigma}{}^{}_\nu \bar{h}^{}_{\rho\sigma}  + \widehat{\nabla}^{}_{(\mu}\mathcal{L}^{}_{\nu)} \nonumber \\ 
& - & \frac{1}{2}\widehat{g}^{}_{\mu\nu}\left(\widehat{\nabla}^{}_{\rho}\mathcal{L}^{\rho}\right)\,,
\label{efes-full}
\end{eqnarray}
where $\widehat{R}^{\rho}_{}{}^{}_\mu{}^{\sigma}{}^{}_\nu$ is the background Riemann tensor and we have used again the background Einstein's field equations [Eq.~\eqref{efes-background}]: $\widehat{G}_{\mu\nu} =0\;\Rightarrow \widehat{R}_{\mu\nu} = \Lambda\widehat{g}_{\mu\nu}$ and $\widehat{R}=\widehat{g}^{\mu\nu}\widehat{R}_{\mu\nu} = 4\Lambda$. Moreover, we have introduced several definitions in Eq.~\eqref{efes-full}. First, we have introduced the trace-reversed metric perturbations:
\begin{equation}
\bar{h}^{}_{\mu\nu} = h^{}_{\mu\nu} - \frac{1}{2}\widehat{g}^{}_{\mu\nu}  h\,,
\end{equation}
with $h$ being the trace of $h_{\mu\nu}$ with respect to the background metric
\begin{equation}
h = \widehat{g}^{\mu\nu}h^{}_{\mu\nu}\,.
\end{equation}
Second, we have introduced the d'Alambertian associated with the background:
\begin{equation}
\widehat{\square}\, \bar{h}^{}_{\mu\nu} = \bar{h}^{}_{\mu\nu;\rho}{}^{;\rho}\,, \qquad
\widehat{\square}\, h = {h}^{}_{;\rho}{}^{;\rho}\,.
\end{equation}
And finally, we have introduced the quantity:
\begin{equation}
\mathcal{L}^{}_{\mu} = \widehat{g}^{\rho\sigma}\widehat{\nabla}^{}_\rho \bar{h}^{}_{\sigma\mu} \,.  
\end{equation}
When we impose $\mathcal{L}^{}_{\mu} = 0$ we are in the so-called Lorenz gauge. But in this paper we are not going to impose any particular gauge, i.e. the developments we present are completely general.

\subsection{Background Solution: Vacuum spherically symmetric spacetimes}

We consider background spacetimes, or regions of spacetimes, that are solutions of the Einstein vacuum equations including a cosmological constant [see Eq.~\eqref{efes-background}].  These solutions come from a generalization of Birkhoff's local uniqueness theorem~\cite{Birkhoff:1923hup} (published before by Jebsen~\cite{Jebsen:1921ori,Jebsen:2005grg}; see also~\cite{Deser:2004gi,VojeJohansen:2005nd}) to the case of a non-vanishing cosmological constant~\cite{Eisland:1925eis} (see~\cite{Schleich:2009uj} for details).  It turns out that the only locally spherically symmetric solutions to Eqs.~\eqref{efes-background} (see~\cite{Schleich:2009uj}) are locally isometric either to one of the Schwarzschild-de Sitter (SchdS~\cite{Kottler:1918}) and Schwarzschild-anti-de Sitter (SchAdS) solutions or to the Nariai spacetime~\cite{Nariai:1950hna,Nariai:1999nar}, which can be seen as the limit of SchdS when the cosmological and event horizons coincide\footnote{We will no consider here the particular case of the Nariai metric as it may require a particular treatment} (see~\cite{Podolsky:1999ts}).  This family of metrics includes very important solutions as the maximally-symmetric solutions of Einstein equations: Minkowski flat spacetime (M; $\Lambda=0$), de Sitter (dS; $\Lambda>0$), and anti-de Sitter (AdS; $\Lambda<0$).
Locally, the background metric can be written in the so-called Schwarzschild form:
\begin{equation}
ds^2=\widehat{g}^{}_{\mu\nu}dx^{\mu}dx^{\nu}=-f(r)\,dt^2+\frac{dr^2}{f(r)}+r^2d\Omega^2\,,
\label{metric}
\end{equation}
where $d\Omega^2 = d\theta^{2}+\sin^{2}\theta d\varphi^{2}$ is the line element of the 2-sphere and $f(r)$ is a function parameterising time translations (related to the redshift).

The solutions described by the metric~\eqref{metric} satisfy Einstein's  equations~\eqref{efes-background}, which become ordinary differential equations (ODEs) for $f(r)$:
\begin{eqnarray}
r f' + f + \Lambda r^2 - 1 = 0 \,, 
\label{efesbackground-1}\\
r \left(f'' + 2 \Lambda \right) + 2 f' = 0 \,.
\label{efesbackground-2}
\end{eqnarray}
There are two combinations of $f(r)$ and its derivatives that are constants and correspond to the cosmological constant $\Lambda$ and the spacetime mass $M$ respectively:
\begin{eqnarray}
\Lambda & = & -\frac{1}{2r^2}\left(r^2 f' \right)' \,, 
\label{exp-cosmological-constant} \\
M & = & \frac{r}{2}\left(1 - f -\frac{\Lambda}{3}r^2 \right) 
   = \frac{r}{2}\left[1 - f + \frac{1}{6}\left(r^2 f'\right)' \right]\,.  
\label{exp-mass}
\end{eqnarray}

In the case of Schwarzschild spacetime~\cite{Schwarzschild:1916uq} (also found independently by Droste~\cite{1917KNAB...19..197D}) we have 
\begin{equation}
f^{}_{\rm Sch} = 1-\frac{r_{s}}{r}\,,
\end{equation}
where $r_s$ is the Schwarzschild radius: $r_{s} = 2GM/c^{2}=2M\,$. In the case of de Sitter and anti-de Sitter spacetimes we have
\begin{equation}
f^{}_{\rm dS} = 1-\frac{r^{2}}{L^2}\,,\qquad
f^{}_{\rm AdS} = 1+\frac{r^2}{L^2}
\end{equation}
where $L$ is the (anti-)de Sitter length scale, which determines the cosmological constant as follows:
\begin{equation}
\Lambda = \pm \frac{3}{L^{2}}\,,   
\label{cosmological-constant}
\end{equation}
where the plus sign corresponds to de Sitter and the minus sign to anti-de Sitter.
Apart from these two cases, we have the case of the Schwarzschild-de Sitter (SchdS) spacetime, which contains the previous two cases in the limits $M\rightarrow 0$ (dS) and $L\rightarrow \infty$ (Sch). This last limit is equivalent to $\Lambda\rightarrow 0$. The function $f(r)$ for Schwarzschild-de Sitter and  -anti de Sitter is:
\begin{equation}
f^{}_{\rm SchdS/SchAdS}(r) = 1-\frac{2M}{r}-\frac{\Lambda}{3}r^2\,,
\end{equation}
where $\Lambda$ is given in Eq.~\eqref{cosmological-constant}.

\subsection{Multipolar expansion of the perturbations of vacuum spherically-symmetric spacetimes} \label{multipolar-expansion}

The background metric can be written as the warped product of two manifolds: $M^2\times_r S^2$, where $M^2$ is a two-dimensional Lorentzian manifold, $r$ is the radial area coordinate, and $S^2$ denotes the $2$-sphere. Therefore, the background metric is given by the semidirect product of a Lorentzian metric on $M^2$, $g_{ab}$, and the unit curvature metric on $S^2$, $\Omega_{AB}$:
\begin{eqnarray}
\widehat{g}^{}_{\mu\nu} = \left(\begin{array}{cc}
g^{}_{ab} & 0 \\[2mm]
0 & r^2\Omega^{}_{AB} \end{array} \right) \,.
\label{warped-metric-background}
\end{eqnarray}
Coordinates on $M^2$ are going to be denoted with lower-case Latin indices, $(x^{a})=(t,r)\,$. Coordinates on $S^2$ are denoted with upper-case Latin indices as: $(\Theta^{A})=(\theta,\varphi)\,$. Then, in connection with Eq.~\eqref{metric} we can write:
\begin{eqnarray}
g^{}_{ab} dx^a dx^b & = & -f(r)\,dt^2+\frac{dr^2}{f(r)} \,, \\
\Omega^{}_{AB} d\Theta^A d\Theta^B & = & d\theta^2 +\sin^2\theta d\varphi^2 \,.
\end{eqnarray}
We use a vertical bar to denotes covariant differentiation on the two-sphere $S^2$ (then $\Omega_{AB|C} = 0$). Similarly, we use a colon to denote covariant differentiation with respect to the metric of the Lorentzian two-dimensional manifold $M^2$, i.e. $g_{ab:c} = 0\,$. On the other hand, the antisymmetric covariant unit tensor associated with the volume form (Levi-Civita tensor) in $S^2$ is denoted by $\epsilon_{AB}$, and the corresponding one on the Lorentzian manifold $M^{2}$ is denoted by $\varepsilon_{ab}\,$. 

The particular geometric structure of the background implies that for certain quantities (e.g. solutions of the wave equation in the background) we can separate the dependence on the coordinates of $M^2$ from the angular dependence, which, in turn, can be expanded in spherical harmonics.
The different harmonics can be divided into even- and odd-parity harmonics depending on how they transform under a parity transformation, $(\theta,\phi)$ $\rightarrow$ $(\pi-\theta, \phi+\pi)$.  If a given harmonic object ${\cal O}^{\ell m}$ transforms as: ${\cal O}^{\ell m} \rightarrow (-1)^{\ell}{\cal O}^{\ell m}$ it is said to be of the even-parity type; while if it transforms as ${\cal O}^{\ell m} \rightarrow (-1)^{\ell+1}{\cal O}^{\ell m}$ it is said to be of the odd-parity type.  With this in mind, the scalar, vector and tensor spherical harmonics are:

~

\noindent $\bullet\,$The scalar harmonics $Y^{\ell m}$ are eigenfunctions of the Laplace operator on the two sphere (see Appendix~\ref{sphericalharmonics}):
\begin{eqnarray}
\Omega^{AB}Y^{\ell m}_{|AB} = -\ell( \ell +1)Y^{\ell m}\,.
\end{eqnarray}

~

\noindent $\bullet\,$The vector spherical harmonics, which are defined for $\ell \geqslant 1$, are given by:
\begin{eqnarray}
Y^{\ell m}_{A} \equiv Y^{\ell m}_{|A} & \quad \text{Even (polar) parity}\,, \\[2mm]
X^{\ell m}_{A} \equiv -\epsilon^{}_A{}^B Y^{\ell m}_{B} &\quad \text{Odd (axial) parity}\,.
\label{vectorharmonics}
\end{eqnarray}

~

\noindent $\bullet\,$The basis of symmetric $2$nd-rank tensor spherical harmonics, which are defined for $\ell \geqslant 2$, are given by 
\begin{eqnarray}
T_{AB}^{\ell m} \equiv Y^{\ell m}\,\Omega_{AB}  \quad & \quad \text{Even parity}\,, \\[2mm]
Y^{\ell m}_{AB} \equiv Y_{|AB}^{\ell m} + \frac{\ell(\ell+1)}{2}Y^{\ell m}\,\Omega_{AB} & \quad \text{Even parity}\,, \\[2mm]
X_{AB}^{\ell m} \equiv X^{\ell m}_{(A|B)}  \quad &\quad \text{Odd parity}\,. 
\label{tensorharmonics}
\end{eqnarray}
Differential properties of these spherical harmonics that are necessary to manipulate the perturbative Einstein equations are given in Appendix~\ref{sphericalharmonics}.

The metric perturbations can be written as a multipole expansion using scalar ($Y^{\ell m}$), vector ($Y^{\ell m}_A$, $X^{\ell m}_A$), and tensor spherical harmonics ($T_{AB}^{\ell m}$, $Y^{\ell m}_{AB}$, $X^{\ell m}_{AB}$).
The main reason for expanding the metric perturbations in this way is that the underlying spherical symmetry prevents different harmonics and different parity modes from mixing, and the perturbation equations can be obtained for each $(\ell,m)$ and parity mode separately (see, e.g.~\cite{Gerlach:1979rw, Gerlach:1980tx}):
\begin{eqnarray}
h_{\mu\nu} = \sum_{\ell ,m} h^{\ell m, \rm odd}_{\mu\nu} + h^{\ell m, \rm even}_{\mu\nu} \,,
\label{eq:2.7}
\end{eqnarray}
where:
\begin{equation}
h^{\ell m, \rm odd}_{\mu\nu}  = \begin{pmatrix} 0 & h_a^{\ell m}X^{\ell m}_A \\[2mm]
  					   		                 \ast & h_2^{\ell m}X^{\ell m}_{AB}  
							    \end{pmatrix}\,, 
\label{metric-perturbation-odd}
\end{equation}
and
\begin{equation}					
h^{\ell m, \rm even}_{\mu\nu}  = \begin{pmatrix} h_{ab}^{\ell m}Y^{\ell m} & \jeven^{\ell m}_aY^{\ell m}_A \\[2mm]
\ast    & r^2\left( K^{\ell m}T^{\ell m}_{AB} + G^{\ell m}Y^{\ell m}_{AB}\right)\end{pmatrix}\,.
\label{metric-perturbation-even}
\end{equation}
Here the asterisk denotes the symmetry on the tensor components, $K^{\ell m}$ and $S^{\ell m}$ denote the scalar perturbations; $h_2^{\ell m}$, and $h^{\ell m}_A$ the vector perturbations; and $h_{ab}^{\ell m}$ the tensorial ones. All of them depend only on the coordinates $\{x^a\}$ of $M^2$.

\subsection{Gauge Invariance}
In relativistic perturbation theory~\cite{Stewart:1974uz,Bruni:1996im} there is a gauge freedom associated with the infinite possible ways of choosing the correspondence between the background and physical spacetimes (see Sec.~\ref{relativistic-perturbation-theory}). In practical terms, this freedom can be associated with the different ways in which we can identify points of the two spacetimes. Taking into account that we can pull back the physical metric into the background tensorial structure [as described by Eq.~\eqref{fundamental-perturbative-equation}], different choices of correspondence between the background and physical spacetimes can be used (from the point of view of the background spacetime) and a coordinate change of the type 
\begin{equation}
x^{\mu}\quad\longrightarrow\quad  x'^{\mu} = {x}^{\mu} + \xi^{\mu}\,,
\label{gauge-transformation}
\end{equation}
where ${x}^{\mu}$ and $x'^{\mu}$ are the coordinates of two points of the physical spacetime, say $p$ and $p'$, that have been identified with a single point of the background spacetime, say $\bar{p}$, by two mappings between the two spacetimes. The mapping between the two points $p$ and $p'$ constitutes what we call a gauge transformation in perturbation theory, and Eq.~\eqref{gauge-transformation} is the coordinate version of such a gauge transformation.  The difference  between the coordinates of the two points $p$ and $p'$ (as seen from the background spacetime)  is described by a vector field, $\xi^{\mu}$, which is the local generator of the gauge transformation, and which is assumed to be small in the same way as we assume that the perturbations are small ($|\xi^{\mu}|\ll|\widehat{g}_{\mu\nu}|$).

The gauge transformation in Eq.~\eqref{gauge-transformation} generates the following transformation of the metric perturbations:
\begin{equation}
h^{}_{\mu\nu}\quad\longrightarrow\quad {h'}^{}_{\mu\nu} = h^{}_{\mu\nu}  -2\,\xi^{}_{(\mu ;\nu)}\,.
\label{gauge-transformation-metric}
\end{equation}

It is important to understand how a general gauge transformation changes the harmonic components of the metric perturbations.  To that end, in the same way we have decomposed the metric perturbations in spherical harmonics we have to do the same with the generator of the gauge transformation $\xi^{\mu}$: For even-parity perturbations, the $(\ell,m)$ harmonic of the gauge generator can be written in the form:
\begin{equation}
\xi^{\ell m, \rm even}_{\mu} dx^{\mu} = \alpha^{\ell m}_{a}(x^{b})dx^{a} + r^{2}\beta^{\ell m}(x^{a})Y^{\ell m}_{A}d\Theta^{A} \,,
\end{equation}
and for odd-parity perturbations
\begin{equation}
\xi^{\ell m, \rm odd}_{\mu} dx^{\mu} =  r^{2}\gamma^{\ell m}(x^{a})X^{\ell m}_{A}d\Theta^{A} \,.
\end{equation}
Note that there are three gauge functions for even-parity perturbation and just one for the odd-parity ones.  

Introducing the multipolar decomposition of the metric perturbations and the gauge vector into Eq.~\eqref{gauge-transformation-metric} we find that the even-parity metric perturbations transform as follows:
\begin{eqnarray}
h'^{\ell m}_{ab} & =  & h^{\ell m}_{ab} - 2\,\alpha^{\ell m}_{(a:b)} \,, \\
\jeven'^{\ell m}_a & = & \jeven^{\ell m}_a - \left(\alpha^{\ell m}_{a} + r^{2}\beta^{\ell m}_{:a} \right)\,, \\
K'^{\ell m} & = & K^{\ell m} + \ell(\ell+1)\beta^{\ell m} -2\frac{r^{:a}}{r}\alpha^{\ell m} \,, \\
G'^{\ell m} & = & G^{\ell m} - 2\beta^{\ell m}\,.
\end{eqnarray}
And the odd-parity metric perturbations transform according to:
\begin{eqnarray}
h'^{\ell m}_{a} & =  & h^{\ell m}_{a} - r^{2}\gamma^{\ell m}_{:a} \,, \\
h'^{\ell m}_{2} & =  & h^{\ell m}_{2} - 2\, r^{2}\gamma^{\ell m} \,.
\end{eqnarray}
There are combinations of the metric perturbations and its derivatives that are invariant under gauge transformations. In the case of even-parity metric perturbations there are four independent gauge-invariant quantities, which can be written as:
\begin{eqnarray}
\tilde{h}^{}_{ab} & = & h^{}_{ab} - \kappa^{}_{a:b} - \kappa^{}_{b:a}\,, 
\label{expression-hathab} \\
\tilde{K}         & = & K + \frac{\ell(\ell+1)}{2} G - 2\frac{r^{a}}{r}\kappa^{}_{a} \,,
\label{expression-hatK}
\end{eqnarray}
where
\begin{eqnarray}
\kappa^{}_{a} & = & \jeven^{}_{a} - \frac{r^{2}}{2}G^{}_{:a}\,, \quad
r^{}_a = r^{}_{:a}~\Rightarrow~r^a=g^{ab}r^{}_b\,.
\end{eqnarray}
In the case of odd-parity metric perturbations there are two independent gauge-invariant quantities:
\begin{equation}
\tilde{h}^{}_{a} = h^{}_{a} -\frac{1}{2}h^{}_{2:a} + \frac{r^{}_{a}}{r} h^{}_{2} \,,
\end{equation}
%

\subsection{Known Master Functions and Equations}\label{master-functions}

Before entering in the search for master functions and equations, let us review the most important known master functions and how they are expressed in covariant form (with respect to the metric $g_{ab}$ of $M^2$; see Sec.~\ref{multipolar-expansion}).  For odd-parity perturbations, the first master function was introduced by Regge and Wheeler~\cite{Regge:1957td} in a pioneering work in black hole perturbation theory. The covariant form of this master function is (see, e.g.~\cite{Martel:2005ir}):
\begin{equation}
\Psi^{}_{\text{RW}} = \frac{r^{a}}{r}\tilde{h}^{}_{a}\,.
\label{regge-wheeler-master-function}
\end{equation}
One can alternatively use the master function introduced by Cunningham, Price, and Moncrief~\cite{Cunningham:1978cp,Cunningham:1979px,Cunningham:1980cp}, which in covariant form reads (see also~\cite{Jhingan:2002kb,Martel:2005ir}):
\begin{equation}
\Psi^{}_{\rm CPM} = \frac{2 r}{(\ell-1)(\ell+2)}\varepsilon^{ab} \left( \tilde{h}^{}_{b:a} -\frac{2}{r}r^{}_{a}\tilde{h}^{}_{b} \right)\,.
\label{cunningham-price-moncrief-master-function}
\end{equation}
A classification of odd-parity master functions can be found in~\cite{Nagar:2005ea}.

In the case of even-parity perturbations we have the master function introduced by Zerilli~\cite{Zerilli:1970la} and later by Moncrief~\cite{Moncrief:1974vm} (see also~\cite{Lousto:1996sx,Martel:2005ir}). It admits the following covariant expression:
\begin{equation}
\Psi^{}_{\rm ZM} = \frac{2 r}{\ell(\ell+1)}\left\{ \tilde{K} + \frac{2}{\lambda}\left(r^{a}r^{b}\tilde{h}^{}_{ab} - r r^{a}\tilde{K}^{}_{:a}\right) \right\}\,,
\label{zerilli-moncrief-master-function}
\end{equation}
where 
\begin{eqnarray}
\lambda(r) & = & r f'-2 \left(f-1\right)+(\ell+2)(\ell-1) \nonumber \\
           & = & (\ell+2)(\ell-1) - \Lambda r^2   -3\left( f-1\right)  \,, 
\label{Lambda-defi} 
\end{eqnarray}
which in the Schwarzschild case reduces to 
\begin{eqnarray}
\lambda(r) & = & (\ell-1)(\ell+2) + \frac{3r^{}_{s}}{r}\,. 
\label{lambda-def-sch}
\end{eqnarray}
All these master functions satisfy wave-type equations in $1+1$ dimensions, with respect to the metric $g_{ab}$ of the Lorentzian manifold $M^2$, with a potential term. The form of these equations in the case of vacuum perturbation looks as follows:
\begin{equation}
\left(\square^{}_2 - \Omega^{}_{\rm even/odd}  \right)\Psi^{}_{\rm even/odd} = 0\,,
\label{master-wave-equation}
\end{equation}
where $\Psi^{}_{\rm even/odd}(t,r)$ is the even/odd master function of choice; $\Omega_{\rm even/odd}(r)$ is the potential, which only depends on the radial area coordinate $r$; and the action of the operator $\square_2$ on any scalar field $\phi$ is given by
\begin{equation}
\square^{}_2\phi  =  g^{ab}\phi^{}_{:ab} \,.
\end{equation}
A slightly different way of introducing the potential comes from the expression of the operator $\square_2$ in Schwarzschild coordinates
\begin{equation}
\square^{}_2\phi  =  -\frac{1}{f}\frac{\partial^{2}\phi}{\partial t^{2}} +\frac{\partial}{\partial r}\left(f\frac{\partial\phi}{\partial r}\right)\,.
\label{expression-box-M2}
\end{equation}
At this point we can introduce {\em tortoise} coordinate:
\begin{equation}
\frac{dr^{}_{\ast}(r)}{dr} = \frac{1}{f(r)}\,.
\label{tortoise-coordinate}
\end{equation}
Combining this with Eq.~\eqref{expression-box-M2} we can write the master equation~\eqref{master-wave-equation} in the following more familiar form:
\begin{equation}
\left(-\frac{\partial^{2}}{\partial t^{2}} + \frac{\partial^{2}}{\partial r^{2}_{\ast}} - V^{}_{\rm even/odd}  \right)\Psi^{}_{\rm even/odd} = 0\,,
\label{coordinate-master-equation}
\end{equation}
where the potential $V_{\rm even/odd}$, is related to the one in Eq.~\eqref{master-wave-equation} by
\begin{equation}
V^{}_{\rm even/odd} = f\,\Omega^{}_{\rm even/odd} \,.
\label{potential-rr*}
\end{equation}
In most places the potential that is used is $V_{\rm even/odd}$, but in this work we will use both. 

To finish this section, we just recall that in the case of the Schwarzschild spacetime, the Regge-Wheeler potential is given by:
\begin{equation}
\Omega^{}_{\rm odd}(r) = \frac{\ell(\ell+1)}{r^{2}} - \frac{3r^{}_{s}}{r^{3}} \,,
\label{schwarzschild-omega-potential-odd-parity}
\end{equation}
while the Zerilli potential is:
\begin{widetext}
\begin{equation}
\Omega^{}_{\rm even}(r) = \frac{1}{\lambda^{2}}\left[ \frac{(\ell-1)^{2}(\ell+2)^{2}}{r^{2}}\left( \ell(\ell+1) + \frac{3r^{}_{s}}{r} \right)  
+ \frac{9r^{2}_{s}}{r^{4}}\left( (\ell-1)(\ell+2) + \frac{r^{}_{s}}{r} \right) \right]\,,
\label{schwarzschild-omega-potential-even-parity}
\end{equation}
\end{widetext}
where $\lambda(r)$ is defined in Eq.~(\ref{Lambda-defi}), and for Schwarzschild is given in Eq.~\eqref{lambda-def-sch}.

\section{Construction of Master Functions and Equations}\label{construction-master-functions}

The main objective of this section, and also of this work, is to look for the most general master function and equation for both odd- and even-parity modes under the following assumptions:
\begin{enumerate}
\item The perturbative gauge is left completely arbitrary. In this way we can check whether or not the master functions have to be necessarily gauge invariant. 
\item The master function is assumed to be linear in the metric perturbations and its first-order derivatives, as it happens for almost all the known master functions. The case of the Bardeen-Press master function~\cite{Bardeen:1973xb} is an exception since it has been derived along the lines of the Teukolsky~\cite{Teukolsky:1972my,Teukolsky:1973ha} procedure to decouple perturbations around Kerr but applied to Schwarzschild. In this case, the decoupling follows from using the Newman-Penrose~\cite{Newman:1961qr} components of the Weyl tensor as master functions. Since the Weyl tensor contains second-order derivatives of the metric perturbations, we cannot recover them from our analysis.
\item The coefficients in the master function are assumed to be time independent. That is, they only depend on the radial area coordinate $r$. This is expected as those coefficients are built from the Lorentzian metric $g_{ab}$ of the 2D manifold $M^2$.
\item The master function satisfies a wave equation of the form~\eqref{master-wave-equation}. The potential is left arbitrary, in the sense that it will be determined only by the perturbative Einstein equations.
\end{enumerate}

In practical terms, the goal is to decouple the equations for the metric perturbations, the perturbative Einstein field equations [Eq.~\eqref{efes-full}].  The perturbation in the Einstein tensor, $\delta G_{\mu\nu} = G_{\mu\nu} -\widehat{G}_{\mu\nu}$, can be expanded in scalar, vector, and tensor spherical harmonics. For a single $(\ell,m)$-harmonic the structure of  $\delta G_{\mu\nu}$ is: 
\begin{widetext}
\begin{eqnarray}
\delta G^{\ell m}_{ab}(x^{c},\Theta^{A}) & = & {\cal E}^{\ell m}_{ab}(x^{c})\;Y^{\ell m}(\Theta^{A})\,, 
\label{PEFEs-ab}\\[2mm]
\delta G^{\ell m}_{aA}(x^{b},\Theta^{B}) & = & {\cal E}^{\ell m}_{a}(x^{b})\;Y^{\ell m}_{A}(\Theta^{B}) + {\cal O}^{\ell m}_{a}(x^{b})\;X^{\ell m}_{A}(\Theta^{B})\,,
\label{PEFES-aA}\\[2mm]
\delta G^{\ell m}_{AB}(x^{a},\Theta^{C}) & = & {\cal E}^{\ell m}_{T}(x^{a})\;T^{\ell m}_{AB}(\Theta^{C}) + {\cal E}^{\ell m}_{Y}(x^{a})\;Y^{\ell m}_{AB}(\Theta^{C}) + {\cal O}^{\ell m}_{X}(x^{a})\;X^{\ell m}_{AB}(\Theta^{C})\,.
\label{PEFES-AB}
\end{eqnarray}
\end{widetext}
We can identify the harmonic components of the perturbative field equations for the even-parity modes, $({\cal E}^{\ell m}_{ab}\,,\,{\cal E}^{\ell m}_{a}\,,\,{\cal E}^{\ell m}_{T}\,,\,{\cal E}^{\ell m}_{Y})$, and for the odd-parity modes, $({\cal O}^{\ell m}_{a}\,,\,{\cal O}^{\ell m}_{X})$.  Their expressions can be constructed in a straightforward way from the expressions of the perturbations of the Ricci tensor given in Appendix~\ref{multipolar-components-perturbations}.

\subsection{Odd-Parity (Axial) Harmonic Modes}\label{master-functions-odd-modes}
In the odd-parity case, we have three independent metric functions, $(h^{\ell m}_{a},h^{\ell m}_{2})$, and the only relevant components of the field equations are ${\cal O}^{\ell m}_{a}$
${\cal O}^{\ell m}_{X}$ [See Eqs.~\eqref{PEFES-aA} and~\eqref{PEFES-AB}].  The most general master function linear (with coefficients depending only on $r$) in the odd-parity metric perturbations and their derivatives is: 
\begin{widetext}
\begin{eqnarray}
\Psi^{\ell m}_{\rm odd}(x^{a}) & = & C^{\ell}_0(r) h^{\ell m}_0(x^{a}) + C^{\ell}_1(r) h^{\ell m}_1(x^{a}) + C^{\ell}_2(r) h^{\ell m}_{2}(x^{a}) + C^{\ell}_3(r) \dot{h}^{\ell m}_0(x^{a}) + C^{\ell}_4(r) h'^{\ell m}_0(x^{a}) + C^{\ell}_5(r) \dot{h}^{\ell m}_1(x^{a})  \nonumber \\
& + & C^{\ell}_6(r) h'^{\ell m}_1(x^{a}) + C^{\ell}_7(r) \dot{h}^{\ell m}_{2}(x^{a}) + C^{\ell}_8(r) h'^{\ell m}_{2}(x^{a}) \,,
\label{ansatz-master-odd}
\end{eqnarray}
\end{widetext}
where we have used the following simplifying notation for time and radial derivatives:
\begin{equation}
\dot{\phi} = \frac{\partial \phi}{\partial t}\,,\qquad 
\phi' = \frac{\partial\phi}{\partial r} \,.
\end{equation}
Notice that the coefficients in Eq.~\eqref{ansatz-master-odd} only depend on the harmonic number $\ell$. Once we have extracted the different harmonics from the Einstein equations, what is left is a set of linear equations in the perturbations $(h^{\ell m}_{a},h^{\ell m}_{2})$ whose structure is (we drop the harmonic indices to simplify the notation):
\begin{eqnarray}
{\cal O}^{}_{t} & : &\quad  \dot{h}'^{}_{1} - h''^{}_{0}  + {\rm LDTs} = 0 \,, 
\label{OtA-eq} \\
{\cal O}^{}_{r} & : &\quad  \ddot{h}^{}_{1} - \dot{h}'^{}_{0} + {\rm LDTs} = 0 \,, 
\label{OrA-eq} \\
{\cal O}^{}_{X} & : &\quad  -\frac{1}{f}\, \ddot{h}^{}_{2} + f\, h''^{}_{2} + {\rm LDTs} = 0 \,,
\label{OAB-eq}
\end{eqnarray}
where LDTs stands for {\em Lower Derivative Terms} (with respect to the other ones), that is, in this particular case they are terms that are linear in the metric perturbations and their first-order derivatives (no second- or higher-order derivatives appear). 
The first step in the analysis of the general solution to Eqs.~\eqref{OtA-eq}-\eqref{OAB-eq} is to study the integrability of Eqs.~\eqref{OtA-eq} and~\eqref{OrA-eq}. Given that they contain $(h''^{}_{0}, \dot{h}'^{}_{1})$ and $(\dot{h}'^{}_{0}, \ddot{h}^{}_{1})$ respectively, we can study their integrability by differentiating Eq.~\eqref{OtA-eq} with respect to $t$ and Eq.~\eqref{OrA-eq} with respect to $r$. It turns out that the integrability condition is identically satisfied by using the three equations [Eqs.~\eqref{OtA-eq}-\eqref{OAB-eq}]. It is important to mention that to arrive to this conclusion we need to use the fact that the background is a solution of Einstein's field equations, which means to use Eqs.~\eqref{efesbackground-1} and~\eqref{efesbackground-2}.

The next step in the analysis is to impose that the function in Eq.~\eqref{ansatz-master-odd} satisfies the following wave equation (assumption 4): 
\begin{equation}
\square^{}_2\Psi^{}_{\rm odd}(x^{a}) = \Omega(r)\, \Psi^{}_{\rm odd}(x^{a}) \,,
\label{key-equation-odd}
\end{equation}
where $\Omega(r)$ is an arbitrary function of $r$ (and $\ell$) that plays the role of the potential [see Eq.~\eqref{master-wave-equation}]. Given the structure of $\Psi^{}_{\rm odd}(x^{a})$ in Eq.~\eqref{ansatz-master-odd}, it is clear that the left-hand side of Eq.~\eqref{key-equation-odd} contains up to third-order derivatives of the metric perturbations $(h^{}_{a},h^{}_{2})$.  In this sense, it is important to realize that Eqs.~\eqref{OtA-eq}-\eqref{OAB-eq} tell us that from the nine possible second-order derivatives of $(h^{}_{a},h^{}_{2})$, three of them can be written in terms of other second-order derivatives and LDTs.  As a consequence, for the twelve possible third-order derivatives of $(h^{}_{a},h^{}_{2})$ we have five independent relations between them\footnote{In principle there should be six (two differentiations of three equations), but the integrability condition between Eqs.~\eqref{OtA-eq} and~\eqref{OrA-eq} eliminates one of them.}. That is, we can write five of the third-order derivatives of $(h^{}_{a},h^{}_{2})$ in terms of the other ones and LDTs, in an independent way.  Therefore, the way to proceed is to expand Eq.~\eqref{key-equation-odd} and use the expressions for the third-order derivatives that we have just mentioned, together with the expressions that relate the second-order derivatives. After we have used all this information, which comes from the perturbative Einstein equations, we just need to impose the vanishing of the coefficients of the metric perturbations $(h^{}_{a},h^{}_{2})$ and their derivatives. That is, once all the possible information coming from Einstein's equations is used, the remainder has to vanish for Eq.~\eqref{key-equation-odd} to hold. Once we have completed this process, we should have obtained the most general odd-parity master function, together with the allowed potential(s).

In our study, the second-order derivatives of the metric perturbations that we are going to eliminate are: $\ddot{h}^{}_{1}\,$, $\dot{h}'^{}_{1}\,$, and $\ddot{h}^{}_{2}\,$. In addition, we also eliminate the third-order derivatives that can be computed from these second-order derivatives, i.e.: $\dddot{h}^{}_{1}\,$, $\ddot{h}'^{}_{1}\,$
$\dot{h}''^{}_{1}\,$, $\dddot{h}^{}_{2}\,$, and $\ddot{h}'^{}_{2}\,$. After eliminating all these derivatives we arrive to an expression of the form:
\begin{widetext}
\begin{eqnarray}
\square^{}_2\Psi^{}_{\rm odd} & = & \tau^{}_{0}\,\dddot{h}^{}_{0} + \tau^{}_{1}\,\ddot{h}'^{}_{0} + \tau^{}_{2}\,\dot{h}''^{}_{0} + \tau^{}_{3}\,{h}'''^{}_{0}
+ \tau^{}_{4}\,{h}'''^{}_{1} \nonumber \\
& + & \tau^{}_{5}\,\ddot{h}^{}_{0} + \tau^{}_{6}\,\dot{h}'^{}_{0} + \tau^{}_{7}\,\dot{h}''^{}_{0} + \tau^{}_{8}\,{h}''^{}_{1} + \tau^{}_{9}\,\dot{h}'^{}_{2} + \tau^{}_{10}\,{h}''^{}_{2} \nonumber \\
& + & \tau^{}_{11}\,\dot{h}^{}_{0} + \tau^{}_{12}\,{h}'^{}_{0} +  \tau^{}_{13}\,\dot{h}^{}_{1} + \tau^{}_{14}\,{h}'^{}_{1} + \tau^{}_{15}\,\dot{h}^{}_{2} + \tau^{}_{16}\,{h}'^{}_{2} 
+ \tau^{}_{17}\,{h}^{}_{0} + \tau^{}_{18}\,{h}^{}_{1} + \tau^{}_{19}\,{h}^{}_{2} \,.
\label{key-equation-expanded}
\end{eqnarray}
\end{widetext}
Notice that there are no third-order derivatives of $h^{}_{2}$. The explanation is that Eq.~\eqref{OAB-eq} can be rewritten as
\begin{equation}
-\frac{1}{2}\square^{}_2 h^{}_{2} + {\rm LDTs} = 0 \,.
\end{equation}
That is, $\square_2 h_{2}$ only produces LDTs and this is why there are no third-order derivatives of $h^{}_{2}$ in Eq.~\eqref{key-equation-expanded}.  Now that only independent derivatives of the metric perturbations appear [Eq.~\eqref{key-equation-expanded}] we can proceed to analyze the consequences of the vanishing of their coefficients. To begin with, the vanishing of $\tau^{}_{0}$ implies
\begin{equation}
C^{}_{3} = 0 \,.
\label{vanishing-C3}
\end{equation}
The vanishing of $\tau^{}_{4}$ leads to
\begin{equation}
C^{}_{6} = 0 \,.
\label{vanishing-C6}
\end{equation}
It turns out that Eq.~\eqref{vanishing-C3} and Eq.~\eqref{vanishing-C6} imply that $\tau^{}_{2}=0$, so there are no extra conditions coming from this term. The vanishing of $\tau^{}_{1}$ and $\tau^{}_{3}$ lead to the same condition
\begin{equation}
C^{}_{5} = -C^{}_{4} \,.
\label{expression-C5}
\end{equation}
This exhausts the information coming from the vanishing of the coefficients of the third-order derivatives of the metric perturbations. Let us now look at the coefficients of the second-order derivatives.  We assume that Eqs.~\eqref{vanishing-C3}-\eqref{expression-C5} hold.  The vanishing of the coefficients $\tau^{}_{6}$ and $\tau^{}_{8}$ yields only one condition (they are equivalent):
\begin{equation}
C^{}_{8} = -\frac{1}{2}C^{}_{1} \,.
\label{expression-C8}
\end{equation}
Introducing this into the equation coming from the vanishing of the coefficient $\tau^{}_{10}$, we obtain an equation for $C^{}_{1}(r)$ 
\begin{equation}
C'^{}_{1}+\left(\frac{1}{r}-\frac{f'}{f}\right)C^{}_{1} =0 ~\Rightarrow~
C^{}_{1}(r) =\frac{K^{}_{1}f(r)}{r} \,,
\end{equation}
where $K^{}_{1}$ is an arbitrary constant. Similarly, the vanishing of the coefficient $\tau^{}_{9}$ implies the following equation for $C^{}_{7}(r)$:
\begin{equation}
r\,C'^{}_{7} + C^{}_{7} = 0 \quad \Rightarrow \quad C^{}_{7}(r) =\frac{K^{}_{7}}{r} \,,
\label{expression-for-C7-odd-parity}
\end{equation}
where $K^{}_{7}$ is another arbitrary constant. The coefficients $\tau^{}_{5}$ and $\tau^{}_{7}$ contain the same information. Their vanishing  allows us to obtain an expression for $C^{}_{4}(r)$:
\begin{equation}
C^{}_{4} = -\frac{r}{2}\left(C^{}_{0}+2\,C^{}_{7}\right) ~\Rightarrow~
C^{}_{4} = -\frac{r}{2}\left(C^{}_{0}+\frac{2\,K^{}_{7}}{r}\right)\,.
\label{expression-for-C4-odd-parity}
\end{equation}
And this exhausts the information coming from the vanishing of the coefficients of the second-order derivatives of the metric perturbations. The analysis of the consequences of the other terms involve the right-hand side of Eq.~\eqref{key-equation-odd}, i.e. the potential.  Then, let us analyze the coefficients of the first-order derivatives. To begin with, since $C^{}_{3}(r)$ vanishes [Eq.~\eqref{vanishing-C3}], the coefficient of $\dot{h}^{}_{0}$, $\tau^{}_{11}$, does not involve $\Omega(r)\,$. It actually provides an expression for $C^{}_{2}(r)$:
\begin{equation}
C^{}_{2}(r) = \frac{C^{}_{1}(r)}{r} = \frac{K^{}_{1}\,f(r)}{r^{2}}\,.
\label{expression-C2}
\end{equation}
The coefficient $\tau^{}_{14}$ vanishes if we introduce this expression for $C^{}_{2}(r)$. From the vanishing of the coefficient $\tau^{}_{15}$ we obtain an expression for the derivative of $C^{}_{0}(r)$ (for $\ell\neq 1$)
\begin{eqnarray}
C'^{}_{0} & = & \frac{2 K^{}_{7}}{(\ell+2)(\ell-1)}\left(\Omega - \frac{f'}{r} \right) \,,
\label{expression-C0prime-old}
\end{eqnarray}
where we have used Eq.~\eqref{efesbackground-2} for the background, i.e. for $f(r)$.
On the other hand, the information in the coefficients $\tau^{}_{12}$ and $\tau^{}_{13}$ is the same.  It is a relationship between the coefficient $C^{}_{0}(r)$ and its first- and second-order derivatives:
\begin{eqnarray}
C''^{}_{0} & = & \left(\frac{2}{r}-\frac{f'}{f} \right)C'^{}_{0} + \frac{\Omega-\Omega^{}_\ast}{f}\left(C^{}_{0} + \frac{2\,K^{}_7}{r}\right) \nonumber \\
& - & \frac{2\,K^{}_7f'}{r^2 f}\,,  
\label{expression-C0primeprime-old} 
\end{eqnarray}
where we have introduced the following definition:
\begin{eqnarray}
\Omega^{}_\ast(r) & = & \Lambda +  \frac{1}{r^{2}}\left[ \ell(\ell+1) + r\left(rf'\right)'+2(f-1)\right]  \nonumber \\
& = & \frac{(\ell+2)(\ell-1) +2 f - r f'}{r^2} \nonumber \\
& = & \Lambda + \frac{\ell(\ell+1) + 3(f-1)}{r^2} \,,
\label{potential-axial}
\end{eqnarray}
where the different equalities appear as a consequence of using the equations for $f(r)$ [Eqs.~\eqref{efesbackground-1} and~\eqref{efesbackground-2}]. By using Eq.~\eqref{expression-C0prime-old} we can eliminate $C'^{}_{0}(r)$ from Eq.~\eqref{expression-C0primeprime-old} and obtain the following alternative expression for $C''^{}_{0}(r)$:
\begin{eqnarray}
C''^{}_{0}(r) & = & \frac{\Omega - \Omega^{}_\ast}{f}C^{}_{0} \nonumber \\
& + & \frac{2K^{}_7 r\Omega^{}_\ast}{(\ell+2)(\ell-1)f}\left( \Omega -  \Omega^{}_\ast  + 2\frac{f-rf'}{r^2}\right)\,. 
\label{expression-C0primeprime}
\end{eqnarray}
Actually, we can also rewrite the equation for $C'^{}_{0}(r)$ [Eq.~\eqref{expression-C0prime-old}] as
\begin{equation}
C'^{}_{0} = \frac{2 K^{}_{7}}{(\ell+2)(\ell-1)}\left(\Omega - \Omega^{}_\ast + o^{}_0 \right)\,,
\label{expression-C0prime}    
\end{equation}
where
\begin{equation}
o^{}_0(r) = \Omega^{}_\ast(r) - \frac{f'(r)}{r} \,.
\label{exp-oo}
\end{equation}
We can integrate the equation for $C'^{}_{0}(r)$ to get:
\begin{widetext}
\begin{equation}
C^{}_{0}(r) = K^{}_0 + \frac{2 K^{}_{7}}{(\ell+2)(\ell-1)}\left\{ \frac{f(r)-1-\Lambda\,r^2}{2\,r} +\int^{}_{r} dr'\Omega(r')\right\} \,,
\label{expression-for-C0-odd-parity}
\end{equation}
\end{widetext}
where $K_0$ is an integration constant.  After all this, the only coefficient of the first-order derivatives of the metric perturbations left to be analyzed is $\tau^{}_{16}$. If we introduce the expression of $C^{}_{2}(r)$ [Eq.~\eqref{expression-C2}] into the coefficient $\tau^{}_{16}$ and impose its vanishing we arrive at the following relation:
\begin{eqnarray}
K^{}_{1}\left[ \Omega(r) - \Omega^{}_\ast(r)  \right] = 0 \,,
\label{bifurcation-point-axial}
\end{eqnarray}
This equation constitutes a bifurcation point in our analysis.  Either $K_1 = 0$ $\Rightarrow$ $C^{}_{1}(r) = 0$ or we have an expression for the potential $\Omega(r)$ in terms of $r$ and $\ell$ [this expression is given in Eq.~\eqref{potential-axial}].  Therefore, there are two branches of possible solutions to the problem we posed at the beginning of this section.

At this point, we only have to focus on the coefficients of the metric perturbations themselves, i.e. $(\tau^{}_{17},\tau^{}_{18},\tau^{}_{19})\,$, and the consistency between the expressions for the first- and second-order derivatives of $C^{}_{0}(r)$ [Eqs.~\eqref{expression-C0prime} and~\eqref{expression-C0primeprime} respectively]. The equation coming from $\tau^{}_{17}$ does not provide new information, in the sense that it becomes a trivial identity $0=0$ if we use the previous information. The coefficients $\tau^{}_{18}$ and $\tau^{}_{19}\,$, after using the previous information and the equations for the background, both lead to Eq.~\eqref{bifurcation-point-axial}.  Therefore, the only thing left is to analyze the compatibility between Eqs.~\eqref{expression-C0prime} and~\eqref{expression-C0primeprime}. This can be done by taking the derivative of $C'^{}_{0}(r)$ by using Eq.~\eqref{expression-C0prime} and subtract $C''^{}_{0}(r)$ from it by using Eq.~\eqref{expression-C0primeprime}.  In the case they were compatible we should be able to reduce the subtraction to an identity $0=0$ by using all the previous information. Otherwise, we should obtain new information/conditions on our unknowns. This is indeed what happens and the new information is encoded in the following equation:
\begin{widetext}
\begin{equation}
\left[\Omega(r) -\Omega^{}_\ast(r)\right]C^{}_{0}(r) + \frac{2K^{}_{7}}{(\ell+2)(\ell-1)}\left[ o^{}_{1}(r)\left( \Omega(r) -\Omega^{}_\ast(r)\right)' + o^{}_{2}(r)\left( \Omega(r) -\Omega^{}_\ast(r)\right) + o^{}_{3}(r)\, \right] = 0\,, 
\label{equation-for-K7-and-C0}
\end{equation}
\end{widetext}
where 
\begin{eqnarray}
o^{}_{1} & = & -f\,, \\
o^{}_{2} & = & r\, \Omega^{}_\ast\,, 
\label{exp-o2}\\
o^{}_{3} & = & 2 \left( \frac{f}{r} - f'\right)\Omega^{}_\ast 
- f\left(\Omega'^{}_\ast -3\frac{f-1}{r^3} - \frac{\Lambda}{r} \right) \,.
\end{eqnarray}
At this point, we have to deal with equations~\eqref{bifurcation-point-axial} and~\eqref{equation-for-K7-and-C0}, taking into account that in this analysis $r$ is arbitrary and $\ell$ is an integer number. Another important point to consider is that $o^{}_{3}(r)$ is in general not zero (it does not vanish everywhere). Therefore, $\Omega(r) = \Omega^{}_\ast(r)$ would imply $K^{}_{7} = 0$. 

On the other hand, Eq.~\eqref{equation-for-K7-and-C0}, in the case $\Omega(r) \neq \Omega^{}_\ast(r)$, can in principle provide an expression for the coefficient $C^{}_{0}(r)$. To that end, we must first make sure that this equation is compatible with the equations for the derivatives of $C_{0}(r)$ [Eqs.~\eqref{expression-C0prime} and~\eqref{expression-C0primeprime}].  Given that equation~\eqref{equation-for-K7-and-C0} is the compatibility between $C'^{}_{0}$ and $C''^{}_{0}$, we just need to check the compatibility with $C'^{}_{0}$ [Eq.~\eqref{expression-C0prime}]. To do so, we can take the derivative of Eq.~\eqref{equation-for-K7-and-C0} and compare it with Eq.~\eqref{expression-C0prime}.  Actually, if we use Eq.~\eqref{expression-C0prime} to eliminate $C'^{}_{0}$ from the derivative of Eq.~\eqref{equation-for-K7-and-C0} we obtain a new relation that has a form very similar to  Eq.~\eqref{equation-for-K7-and-C0}:
\begin{equation}
\Gamma[r,\Omega]\, C^{}_{0}(r) + K^{}_{7}\,\Delta[r,\Omega] = 0\,,  
\label{general-compatibility-relation-axial}
\end{equation}
where $\Gamma[r,\Omega]$ and $\Delta[r,\Omega]$ are functionals of $\Omega(r)$. 
From these two relations [Eqs.~\eqref{equation-for-K7-and-C0} and~\eqref{general-compatibility-relation-axial}] we can eliminate either $C^{}_{0}(r)$ or $K_7$. In any case, and assuming that $\Omega(r) \neq \Omega^{}_\ast(r)$, we obtain an equation for $\Omega''(r)$ that has the following form:
\begin{widetext}
\begin{equation}
K^{}_7\left[ \left( \hat{o}^{}_{1}(r) \frac{\delta\Omega'(r)}{\delta\Omega(r)}\right)' + \left(\frac{\hat{o}^{}_{3}(r) }{\delta\Omega(r)}\right)' + \hat{o}^{}_{o}(r) + \hat{o}'^{}_{2}(r) - \delta\Omega(r) \right] = 0 \,,
\label{equation-for-omega-primeprime}
\end{equation}
\end{widetext}
where we have introduced the following definitions:
\begin{eqnarray}
\delta\Omega(r) & = & \Omega(r) -\Omega^{}_\ast(r)\,, \\
\hat{o}^{}_{I}(r) & = & - o^{}_{I}(r)\qquad (I=0-3) \,.
\label{exp-hat-oI}
\end{eqnarray}  
In this way, $\hat{o}^{}_{1}(r)=f(r)$. From Eq.~\eqref{equation-for-omega-primeprime}, it is clear that if $K^{}_7\neq 0$, we have an equation for  $\Omega''(r)$ which is non-linear in $\Omega(r)$.

In principle, we can differentiate Eq.~\eqref{general-compatibility-relation-axial} and we would get an expression of the same form by using again Eq.~\eqref{expression-C0prime} for $C'^{}_{0}(r)$. Combining the new equation with Eq.~\eqref{equation-for-K7-and-C0} we can again eliminate either $C^{}_{0}(r)$ or $K_7$. This would provide us with a new equation where the only unknown is the potential $\Omega(r)$. It turns out that using the equation for  $\Omega''(r)$ provided by Eq.~\eqref{equation-for-omega-primeprime}, we get an identity $0=0$, and this ends the chain of possible equations of the form in Eq.~\eqref{general-compatibility-relation-axial}.  

In summary, we end up with two different branches. The first branch is determined by the following relation
\begin{equation}
\Omega(r) = \Omega^{}_\ast(r) \quad\Longrightarrow\quad \delta\Omega(r) = 0\,.
\end{equation}
Then, the potential for odd-parity perturbations, $\Omega^{}_\ast(r)$, is given by Eq.~\eqref{potential-axial}. In the case of a Schwarzschild background, this potential has been shown in Eq.~\eqref{schwarzschild-omega-potential-odd-parity}, while for the case of a de Sitter background, it is simply the centrifugal barrier (the same one as in a Minkowski, and also anti-de Sitter, background):
\begin{equation}
\Omega^{\rm dS}_{\ast}(r) = \frac{\ell(\ell+1)}{r^2}\,.
\label{centrifugal-barrier-potential}
\end{equation}
In this first branch, Eq.~\eqref{bifurcation-point-axial} is automatically satisfied. On the other hand, Eq.~\eqref{equation-for-K7-and-C0} implies:
\begin{equation}
K^{}_7 = 0 \,.
\end{equation}
Therefore, according to Eq.~\eqref{expression-C0prime}, the coefficient $C^{}_{0}(r)$ is constant:
\begin{equation}
C''^{}_{0}(r) = C'^{}_{0}(r) = 0\quad\Longrightarrow\quad C^{}_{0}(r) = K^{}_0\,,
\end{equation}
where $K^{}_0$ is a constant. This is also compatible with the equation for $C''^{}_{0}(r)$ [Eq.~\eqref{expression-C0primeprime}]. Then, the most general odd-parity master function in this branch is
\begin{widetext}
\begin{equation}
\Psi^{}_{\rm odd}(t,r) =\frac{rK^{}_0}{2} \left(\dot{h}^{}_1(t,r) - h'^{}_0(t,r)\right) + K^{}_0 h_0(t,r) 
+ K^{}_1\frac{f(r)}{r} h^{}_1(t,r) + \frac{K^{}_1f(r)}{r}\left(  \frac{1}{r} h^{}_2(t,r) -\frac{1}{2} h'^{}_2(t,r) \right)\,.  
\end{equation}
\end{widetext}
Using the definitions introduced in Sec.~\ref{master-functions} we can rewrite this master function in a completely covariant way:
\begin{equation}
\Psi^{}_{\rm odd}(t,r) = K^{}_1\,\Psi^{}_{\rm RW}(t,r) - \frac{(\ell+2)(\ell-1)}{4} K^{}_0\,\Psi^{}_{\rm CPM}(t,r) \,.
\end{equation}
That is, the most general master function in the first branch is a linear combination of the Regge-Wheeler and Cunningham-Price-Moncrief master functions [see Eqs.~\eqref{regge-wheeler-master-function} and~\eqref{cunningham-price-moncrief-master-function} respectively]. Therefore, it is covariant and gauge-invariant by construction. It is important here to note that the Regge-Wheeler master function turns out to be the time derivative of the Cunningham-Price-Moncrief master function~\cite{Martel:2005ir}: 
\begin{equation}
t^a \Psi^{}_{{\rm CPM}\,:a} = 2\,\Psi^{}_{\rm RW}\,. 
\label{CPMdot-RW}
\end{equation}
Finally, the potential corresponds to the already known potential, namely the Regge-Wheeler one. This ends the analysis of the first branch.  

Let us now consider the second branch. It is characterized by $\Omega(r) \neq \Omega^{}_\ast(r)$ (i.e. $\delta\Omega(r)\neq 0$) and hence, we must necessarily have
\begin{equation}
K^{}_1 = 0 \,,
\end{equation}
and $\delta\Omega(r)$ has to satisfy Eq.~\eqref{equation-for-omega-primeprime}.  In this case, the only non-zero coefficients of $\Psi^{}_{\rm odd}$ are: $C_{0}(r)$ [Eq.~\eqref{expression-for-C0-odd-parity}],
$C_{4}(r)$ [Eq.~\eqref{expression-for-C4-odd-parity}], $C_{5}(r)$ [$=-C_{4}(r)$], and $C_{7}(r)$ [Eq.~\eqref{expression-for-C7-odd-parity}]. Introducing these expressions we can write the most general master function in the second branch as
\begin{widetext}
\begin{equation}
\Psi^{}_{\rm odd}(t,r) = - \frac{(\ell+2)(\ell-1)}{4} \left( K^{}_0 + \hat{K}^{}_7\,\Xi(r) \right) \Psi^{}_{\rm CPM}(t,r)  + {K}^{}_7\,\Phi^{}_{\rm ON}(t,r) \,,
\label{master-function-odd-second-branch}
\end{equation}
\end{widetext}
where:
\begin{eqnarray}
& & \hat{K}^{}_7 = \frac{2 K^{}_{7}}{(\ell+2)(\ell-1)} \,, \\
& & \Xi(r) = \frac{f(r)-1-\Lambda\,r^2}{2\,r} +\int^{}_r dr'\Omega(r') \,, \\
\end{eqnarray}
and $\Phi^{}_{\rm ON}(t,r)$ is a new odd-parity function that can be given in a completely covariant form as
\begin{eqnarray}
\Phi^{}_{\rm ON}(t,r) = \varepsilon^{ab} \tilde{h}^{}_{a:b} \,.
\label{odd-new-function}
\end{eqnarray}
One can check that $\Phi^{}_{\rm ON}(t,r)$ is also a gauge-invariant quantity although it is not by itself a master function.  However, the combination with $\Psi^{}_{\rm CPM}(t,r)$ that appears in Eq.~\eqref{master-function-odd-second-branch}, whose coefficient is $K_7$, is an odd-parity master function.  Then, the most general odd-parity master function in the second branch, Eq.~\eqref{master-function-odd-second-branch}, is fully covariant and gauge invariant. The potential in this second branch is any function satisfying the non-linear ODE of Eq.~\eqref{equation-for-omega-primeprime}. Regarding this equation, it is worth noting that if write it in terms of $\delta V = f\delta\Omega$ [see Eq.~\eqref{potential-rr*}], use the expressions for $\hat{o}_o$ and $\hat{o}_2$, which satisfy
\begin{equation}
\hat{o}_o + \hat{o}'^{}_2 - f'' = 0 \,.
\end{equation}
and exchange derivatives with respect to $r$ with derivatives with respect to the tortoise coordinate we arrive at the following simpler equation:
\begin{equation}
\left(\frac{\delta{V}^{}_{,x}}{\delta{V}} \right)^{}_{,x}
+ 2 \left(\frac{V^{\rm odd}_{,x}}{\delta{V}} \right)^{}_{,x} - \delta{V} = 0 \,,
\label{xdarboux-odd}
\end{equation}
where $V^{\rm odd}_{} = f\Omega_{\ast}$ is the Regge-Wheeler potential. 
Finally, it is important to remark that changing the potential we are changing at the same time the master function. This ends the analysis of the odd-parity case.

\subsection{Even-Parity (Polar) Harmonic Modes}\label{master-functions-polar-modes}

In the even-parity case (polar perturbations) we have seven independent metric functions, $(h^{\ell m}_{ab},\jeven^{\ell m}_{a},K^{\ell m},G^{\ell m})$, and this time we have seven relevant field equations, coming from the components $({\cal E}^{\ell m}_{ab}\,,\,{\cal E}^{\ell m}_{a}\,,\,{\cal E}^{\ell m}_{T}\,,\,{\cal E}^{\ell m}_{Y})$ of the perturbative field equations.  Taking into account the assumptions we imposed on the master function before, the most general ansatz to start with is~\footnote{As in the odd-parity case, we use coefficients named $C^{\ell}_I$ but no confusion should arise since they are purely auxiliary quantities and there are no cross references.}
\begin{widetext}
\begin{eqnarray}
\Psi^{\ell m}_{\rm even} & = & C^{\ell}_0\; h^{\ell m}_{00} + C^{\ell}_1\; h^{\ell m}_{01} + C^{\ell}_2\; h^{\ell m}_{11} + C^{\ell}_3\; \jeven^{\ell m}_0 + C^{\ell}_4\; \jeven^{\ell m}_1 + C^{\ell}_5\; K^{\ell m} + C^{\ell}_6\; G^{\ell m} \nonumber \\
& + & C^{\ell}_7\; \dot{h}^{\ell m}_{00} + C^{\ell}_8\; h'^{\ell m}_{00} + C^{\ell}_9\; \dot{h}^{\ell m}_{01} + C^{\ell}_{10}\; h'^{\ell m}_{01}  + C^{\ell}_{11}\; \dot{h}^{\ell m}_{11} + C^{\ell}_{12}\; h'^{\ell m}_{11} \nonumber \\
& + & C^{\ell}_{13}\; \dot{\jeven}^{\ell m}_0 + C^{\ell}_{14}\; \jeven'^{\ell m}_0 + C^{\ell}_{15}\; \dot{\jeven}^{\ell m}_1 + C^{\ell}_{16}\; \jeven'^{\ell m}_1 + C^{\ell}_{17}\; \dot{K}^{\ell m} + C^{\ell}_{18}\; K'^{\ell m} + C^{\ell}_{19}\; \dot{G}^{\ell m} + C^{\ell}_{20}\; G'^{\ell m} \,.
\label{ansatz-master-even}
\end{eqnarray}
\end{widetext}
For the sake of simplicity we have hidden the dependence of the different functions since it is clear that the coefficients $C^{\ell}_I$ ($I=0,\ldots,20$) only depend on the radial coordinate $r$, and the metric perturbations $(h^{\ell m}_{ab},\jeven^{\ell m}_{a},K^{\ell m},G^{\ell m})$ depend on the coordinates of $M^{2}$, i.e. $\{x^{a}\}\,$.  Following the procedure of the odd-parity case, let us analyze the structure of the perturbative field equations for the metric perturbations $(h^{\ell m}_{ab},\jeven^{\ell m}_{a},K^{\ell m},G^{\ell m})\,$  This essentially means to analyze the structure of the terms containing second-order derivatives.  Dropping again the harmonic indices, we find that the equations of interest have the following form: 
\begin{eqnarray}
{\cal E}^{}_{tt} & : &~ K'' + {\rm LDTs} = 0 \,, 
\label{Ett-eq} \\
{\cal E}^{}_{tr} & : &~ \dot{K}' + {\rm LDTs} = 0 \,, 
\label{Etr-eq} \\
{\cal E}^{}_{rr} & : &~ \ddot{K} + {\rm LDTs} = 0 \,, 
\label{Err-eq} \\
{\cal E}^{}_{tY} & : &~ \dot{\jeven}'^{}_{1} - \jeven''^{}_{0} + {\rm LDTs} = 0 \,, 
\label{EtY-eq} \\
{\cal E}^{}_{rY} & : &~ \ddot{\jeven}^{}_{1} - \dot{\jeven}'^{}_{1} + {\rm LDTs} = 0 \,, 
\label{ErY-eq} \\
{\cal E}^{}_{T}  & : &  -\frac{1}{f}\, \ddot{K} + f\, K'' - h''^{}_{00} + 2\dot{h}'^{}_{01} -\ddot{h}^{}_{11} \nonumber \\
                 &  & +\, {\rm LDTs} = 0 \,, 
\label{ETT-eq} \\
{\cal E}^{}_{Y}  & : &  -\frac{1}{f}\, \ddot{G} + f\, G'' + {\rm LDTs} = 0 \,.
\label{EYY-eq}
\end{eqnarray}
Given that Eqs.~\eqref{EtY-eq} and~\eqref{ErY-eq} contain $(\jeven''^{}_{0}, \dot{\jeven}'^{}_{1})$ and $(\dot{\jeven}'^{}_{0}, \ddot{\jeven}^{}_{1})$ respectively, we can study their integrability by differentiating Eq.~\eqref{EtY-eq} with respect to $t$ and Eq.~\eqref{ErY-eq} with respect to $r$. The result is that this integrability condition is identically satisfied by using the the following equations: \eqref{Ett-eq}, \eqref{Err-eq}, \eqref{ErY-eq}, \eqref{ETT-eq}, and~\eqref{EYY-eq}. 
Like in the odd-parity case, in order to arrive to this conclusion we have used the equations for the background metric function $f(r)$. Similarly, Eqs.~\eqref{Ett-eq}-\eqref{Err-eq} contain all the second-order derivatives of $K$, i.e.~$(\ddot{K}\,,\,\dot{K}'\,,\,K'')$. 
One can show that their integrability conditions are satisfied by using the other equations and that the background is a solution of Einstein's equations. The fact that the integrability conditions are satisfied, both for odd- and even-parity perturbations, is intimately related with the 
metric perturbations satisfying a linearized version of the (contracted) second Bianchi identities: $\bar{g}^{\rho\mu}\bar{\nabla}_{\rho}\delta G_{\mu\nu}=0$. On the other hand, we can substitute the expressions for the second-order derivatives of $K$ [Eqs.~\eqref{Ett-eq}-\eqref{Err-eq}] into Eq.~\eqref{ETT-eq} so that it becomes a relation between second-order derivatives of $h^{}_{ab}$.

Like in the odd-parity case, we impose our master function candidate in Eq.~\eqref{ansatz-master-even} to satisfy a wave equation of the type
\begin{equation}
\square^{}_2\Psi^{}_{\rm even}(x^{a}) = \Omega(r)\, \Psi^{}_{\rm even}(x^{a}) \,,
\label{key-equation-even}
\end{equation}
where $\Omega(r)$ is a function of $r$ (and $\ell$) to be determined and that will play the role of the potential for the dynamics of even-parity perturbations. When we insert the general even-parity master function of Eq.~\eqref{ansatz-master-even} we will get again a linear combination of the metric perturbations $(h^{\ell m}_{ab},\jeven^{\ell m}_{a},K^{\ell m},G^{\ell m})$  and their derivatives up to third-order.  However, not all these derivatives are independent since Eqs.~\eqref{Ett-eq}-\eqref{EYY-eq} already determine a subset of second-order derivatives (seven of them), and hence a subset of the third-order derivatives too.  To be more specify, let us make a choice: From the perturbative Einstein equations we determine the following second-order derivatives: $\ddot{h}^{}_{11}\,$, $\ddot{\jeven}^{}_{1}\,$, $\dot{\jeven}'^{}_{1}\,$, $\ddot{K}\,$, $\dot{K}'\,$, $K''\,$, and $\ddot{G}\,$. We can then substitute these derivatives into Eq.~\eqref{key-equation-even}.  In addition, we can also substitute the third-order derivatives that can be estimated from them, namely: $\dddot{h}^{}_{11}\,$, $\ddot{h}'^{}_{11}\,$, $\dddot{\jeven}^{}_{1}\,$, $\ddot{\jeven}'^{}_{1}\,$, $\dot{\jeven}''^{}_{1}\,$, $\dddot{K}\,$, $\ddot{K}'\,$, $\dot{K}''\,$, $K'''\,$, $\dddot{G}\,$, and $\ddot{G}'\,$.  Once this is done we arrive to an expression of the form\footnote{As in the odd-parity case, we use coefficients named $\tau_I$ but no confusion should arise since they are purely auxiliary quantities and there are no cross references.}:
\begin{widetext}
\begin{eqnarray}
\square^{}_2\Psi^{}_{\rm even} & = & \tau^{}_{0}\,\dddot{h}^{}_{00} + \tau^{}_{1}\,\ddot{h}'^{}_{00} + \tau^{}_{2}\,\dot{h}''^{}_{00} + \tau^{}_{3}\,{h}'''^{}_{00}
+ \tau^{}_{4}\,\dddot{h}^{}_{01} + \tau^{}_{5}\,\ddot{h}'^{}_{01} + \tau^{}_{6}\,\dot{h}''^{}_{01} + \tau^{}_{7}\,{h}'''^{}_{01} 
+ \tau^{}_{8}\,\dot{h}''^{}_{11} + \tau^{}_{9}\,{h}'''^{}_{11} \nonumber \\
& + & \tau^{}_{10}\,\dddot{\jeven}^{}_{0} + \tau^{}_{11}\,\ddot{\jeven}'^{}_{0} + \tau^{}_{12}\,\dot{\jeven}''^{}_{0} + \tau^{}_{13}\,{\jeven}'''^{}_{0}
+ \tau^{}_{14}\,{\jeven}'''^{}_{1}\nonumber \\
& + & \tau^{}_{15}\,\ddot{h}^{}_{00} + \tau^{}_{16}\,\dot{h}'^{}_{00} + \tau^{}_{17}\,\dot{h}''^{}_{00} + \tau^{}_{18}\,\ddot{h}^{}_{01} + \tau^{}_{19}\,\dot{h}'^{}_{01} + \tau^{}_{20}\,\dot{h}''^{}_{01} + \tau^{}_{21}\,\dot{h}'^{}_{11} + \tau^{}_{22}\,\dot{h}''^{}_{11} \nonumber \\
& + & \tau^{}_{23}\,\ddot{\jeven}^{}_{0} + \tau^{}_{24}\,\dot{\jeven}'^{}_{0} + \tau^{}_{25}\,\dot{\jeven}''^{}_{0} + \tau^{}_{26}\,\dot{\jeven}''^{}_{1} + \tau^{}_{27}\,\dot{G}'^{}_{0} + \tau^{}_{28}\,G''^{}_{0} \nonumber \\
& + & \tau^{}_{29}\,\dot{h}^{}_{00} + \tau^{}_{30}\,{h}'^{}_{00} +  \tau^{}_{31}\,\dot{h}^{}_{01} + \tau^{}_{32}\,{h}'^{}_{01} + \tau^{}_{33}\,\dot{h}^{}_{11} + \tau^{}_{34}\,{h}'^{}_{11} +  \tau^{}_{35}\,\dot{\jeven}^{}_{0} + \tau^{}_{36}\,{\jeven}'^{}_{0} + \tau^{}_{37}\,\dot{\jeven}^{}_{1} + \tau^{}_{38}\,{\jeven}'^{}_{1} \nonumber\\
& + & \tau^{}_{39}\,\dot{K} + \tau^{}_{40}\,{K}' + \tau^{}_{41}\,\dot{G} + \tau^{}_{42}\,{G}' \nonumber\\
& + & \tau^{}_{43}\,{h}^{}_{00} + \tau^{}_{44}\,{h}^{}_{01} + \tau^{}_{45}\,{h}^{}_{11} + \tau^{}_{46}\,{\jeven}^{}_{0} + \tau^{}_{47}\,{\jeven}^{}_{1} + \tau^{}_{48}\,K + \tau^{}_{49}\,G \,.
\label{key-equation-expanded-even}
\end{eqnarray}
\end{widetext}
As it already happened in the odd-parity case with the metric perturbation $h^{}_{2}$, here there are no third-order derivatives of $G$, and it is due to the same reason. If we look at Eq.~\eqref{EYY-eq}, it turns out it can be rewritten, up to a numerical factor, as $\square_2 G + {\rm LDTs} = 0\,$. And since the operator $\square_2$ only produces LDTs, no third derivatives of $G$ appear in Eq.~\eqref{key-equation-expanded-even}.

Let us now analyze the implications of the vanishing of the coefficients $\tau_{I}$ ($I=0\,,\,\ldots\,,49$) for the general form of the even-parity master function. To begin with, the vanishing of the coefficients $\tau_{0},\ldots,\tau_{5}$ implies the vanishing of the following coefficients of the master function [see Eq.~\eqref{ansatz-master-even}]:
\begin{equation}
C^{}_{7} = C^{}_{8} = C^{}_{9} = C^{}_{10} = C^{}_{11} = C^{}_{12} = 0\,.
\end{equation}
We have arrived at this result by using the information from one coefficient into the next one. Notice that this implies that the master function cannot contain any first-order derivatives of the metric perturbations $h^{}_{ab}$. 
Using this information, the vanishing of the coefficients $\tau_{6}\,,\ldots\,,\tau_{9}$ does not provide any additional information. The vanishing of $\tau_{10}\,$ and $\tau_{12}\,$ imply:
\begin{equation}
C^{}_{13} = C^{}_{16} = 0\,.
\end{equation}
Taking this into account, the coefficient $\tau^{}_{14}$ does not provide any additional information. The coefficients $\tau_{11}$ and $\tau_{13}$ contain the same information:
\begin{equation}
C^{}_{14} + C^{}_{15} = 0\,.
\end{equation}
The vanishing of the coefficient $\tau^{}_{15}$ tells us that
\begin{equation}
C^{}_{0} = 0\,.
\end{equation}
The coefficient $\tau^{}_{16}$, together with the previous information, leads to
\begin{equation}
C^{}_{17} = \frac{r}{f}\,C^{}_{14}\,.
\end{equation}
Similarly, from the coefficient $\tau^{}_{17}$ we have
\begin{equation}
C^{}_{18} = -\frac{r}{f}\,C^{}_2\,. 
\label{expression-for-C18-even}
\end{equation}
From $\tau^{}_{18}$ we get
\begin{equation}
C^{}_{14} = -C^{}_1\,.
\end{equation}
Using all the information obtained up to now, we can see that the coefficients $\tau_{19}\,,\ldots\,,\tau_{22}$ do not provide new information. Instead, from $\tau_{23}$ we get an expression for $C_{19}$ in terms of $C_{1}$ and $C_{3}$:
\begin{equation}
C^{}_{19} = \frac{2\,f-\ell(\ell+1)}{2\,f}\,r\,C^{}_1 - \frac{r^2}{2}C^{}_3 \,.
\label{expC19-even}
\end{equation}
From $\tau_{24}$ we obtain a similar expression for $C_{20}$
\begin{equation}
C^{}_{20} = -\frac{\ell(\ell+1)}{2\,f}\,r\,C^{}_2 - \frac{r^2}{2}C^{}_4 \,.
\end{equation}
With this, the conditions coming from $\tau_{25}$ and $\tau_{26}$ are automatically satisfied. The coefficient $\tau_{27}$ leads to a simple ODE: 
\begin{equation}
r C'^{}_{19}-C^{}_{19}=0\,. 
\label{odeC19}
\end{equation}
Taking into account Eq.~\eqref{expC19-even} we can solve the equation for $C_{3}$:
\begin{equation}
C^{}_3 = \frac{K^{}_{13}}{r} + \frac{2\,f - \ell(\ell+1)}{r\, f} C^{}_1 \,,
\end{equation}
where $K_{13}$ is an integration constant. After substitution in Eq.~\eqref{expC19-even}, the coefficient $C_{19}$ takes the following simple form [compatible with Eqs.~\eqref{expC19-even} and~\eqref{odeC19}]:
\begin{equation}
C^{}_{19}(r) = -\frac{K^{}_{13}\,r}{2}\,.
\end{equation}
Similarly, from $\tau_{28}$ we can obtain an ODE from which we find an expression for $C_{4}$
\begin{equation}
C^{}_4 = K^{}_{24}\,\frac{f}{r} - \frac{\ell(\ell+1)}{r\, f} C^{}_2 \,,
\label{expression-for-C4-even}
\end{equation}
where $K_{24}$ is another integration constant. And thanks to this, the coefficient $C_{20}$ takes the following simple form:
\begin{equation}
C^{}_{20}(r) = -\frac{K^{}_{24}}{2}\,r f(r) \,. 
\label{expression-for-C20-even}
\end{equation}
Using the previous expression, the condition coming from $\tau_{29}$ determines the coefficient $C_{1}$: 
\begin{equation}
C^{}_1 = \frac{K^{}_{13}\,f}{\ell(\ell+1) + r f' - 2\,f} \,,
\label{expC1-even}
\end{equation}
and the coefficient $\tau_{30}$ determines the coefficient $C_{5}$:
\begin{equation}
C^{}_{5} = -\frac{K^{}_{24}}{2} + \frac{(\ell+2)(\ell-1)+3\,rf'+2\Lambda r^{2}}{2\,f^{2}}C^{}_{2}\,.
\label{expression-for-C5-even}
\end{equation}
With all this information the relations coming from the coefficients $\tau_{31}\,,\ldots\,,\tau_{34}$ are satisfied. The coefficient $\tau_{35}$ provides an
expression for $C^{}_{6}$:
\begin{equation}
C^{}_{6} = \frac{\ell(\ell+1)}{4}\left( \frac{(\ell+2)(\ell-1)+3\,rf'+2\Lambda r^{2}}{f^{2}}C^{}_{2} - K^{}_{24}   \right)\,.
\label{expression-for-C6-even}
\end{equation}
To sum up the situation until now: All the non-zero coefficients are ultimately found either in terms of $C_{1}$ or $C_{2}$. The coefficient $C_{1}$ has already been determined in terms of $r$, $f$, and an integration constant, $K_{13}$ [see Eq.~\eqref{expC1-even}]. Only $C_{2}$ and the potential $\Omega$ have to be determined from the equations imposed by the rest of coefficients $\tau_{I}\,$.  In what follows we use all the information found until now.

The coefficient $\tau_{38}$ brings no new information. The equation coming from the coefficient $\tau_{36}$ can be written as follows:
\begin{equation}
K^{}_{13}\left[\Omega(r)-\Omega^{}_{\ast}(r)\right] = 0\,,
\label{exptau36}
\end{equation}
where $\Omega^{}_{\ast}(r)$ is a given function of $r$, $f$ (and its derivatives), $\ell$, and $\Lambda$.  As in the odd-parity case [see Eq.~\eqref{bifurcation-point-axial}], this equation is also a branch point in the analysis. By doing some algebra (and using the field equations of the background spacetime) we can write $\Omega_{\ast}(r)$ as follows:
\begin{widetext}
\begin{equation}
\Omega^{}_{\ast}(r) = \frac{ \lambda^3(r) - 2 \Lambda r^2\left[\lambda(r) - (\ell+2)(\ell-1)\right]^2 +2\left(\ell+2\right)^2\left(\ell-1\right)^2 \left(\ell^2+\ell+1\right)}{3 r^2 \lambda^2(r)}\,,
\label{potential-even}
\end{equation}
\end{widetext}
where $\lambda(r)$ was defined in Eq.~\eqref{Lambda-defi}. The relations coming from the vanishing of the coefficients $\tau_{37}\,$, $\tau_{39}\,$, $\tau_{41}\,$, $\tau_{44}\,$, and $\tau_{46}$ also reduce to the bifurcation point represented by Eq.~\eqref{exptau36}. To get there we have used sometimes the equations for the background metric [Eqs.~\eqref{efesbackground-1} and~\eqref{efesbackground-2}]. On the other hand, from the vanishing of the coefficient $\tau_{40}$ we arrive at an expression for the second-order derivative of the coefficient $C_{2}$:
\begin{equation}
C''^{}_{2} = e^{}_{1}\,C'^{}_{2} + e^{}_{2}\,C^{}_{2} +  K^{}_{24}\frac{f}{r^{2}} \,,
\label{expd2C2-even}
\end{equation}
where the coefficients $e^{}_{1}(r)$ and $e^{}_{2}(r)$, using the field equations of the background [Eqs.~\eqref{efesbackground-1} and~\eqref{efesbackground-2}], are
\begin{eqnarray}
e^{}_{1} & = & \frac{\ell(\ell+1) -3(\Lambda r^2-1) - f}{r f} \,, \\
e^{}_{2} & = & \frac{f^2 + (1-5\Lambda r^2)f + 2(r^2\Lambda-1)(\ell^2+\ell+2-2\Lambda r^2)}{r^2f^2} \nonumber \\
         & + & \frac{\Omega}{f}\,.
\end{eqnarray}
As we can see, the function $e^{}_{2}(r)$ contains the potential $\Omega(r)$. On the other hand, from the coefficient $\tau_{42}$ we arrive at an expression for the first derivative of the coefficient $C_{2}$ of the form
\begin{equation}
C'^{}_{2} = e^{}_{4}\, C^{}_{2} + K^{}_{24}\,e^{}_{3} \,, 
\label{expdC2-even}
\end{equation}
where the coefficients $e_3(r)$ and $e_4(r)$ are
\begin{eqnarray}
e^{}_{3} & = & -f^2 \;\frac{r^2 \Omega + \lambda -(\ell+2)(\ell-1)}{\ell \left(\ell+1\right) r \lambda} \,, \\
e^{}_{4} & = &  \left(\ln\frac{r f^2}{\lambda} \right)'\,.
\label{coefficient-e4-even}
\end{eqnarray}
Here, the function $e^{}_{3}(r)$ is the one that contains the potential $\Omega(r)$.  It turns our that Eq.~\eqref{expdC2-even} can be integrated to obtain the following expression for $C_2(r)$:
\begin{widetext}
\begin{eqnarray}
C^{}_{2}(r) = \frac{r f^2(r)}{\lambda(r)} \left\{ K^{}_2 + \frac{K^{}_{24}}{\ell(\ell+1)}\left[ \frac{\lambda(r)-(\ell+2)(\ell-1)}{2 r} - \int^{}_r dr'\,\Omega(r')\right] \right\} \,.
\label{expression-CE2}
\end{eqnarray}
\end{widetext}
where $K_2$ is another integration constant.

Going back to the analysis of the $\tau_I$ coefficients, using previous information [including the equations for $C'^{}_{2}$, Eq.~\eqref{expdC2-even}, and $C''^{}_{2}$, Eq.~\eqref{expd2C2-even}] and some algebra, it is possible to see that the equations coming from the vanishing of the coefficients $\tau_{43}\,$, $\tau_{45}\,$, $\tau_{47}\,$, $\tau_{48}\,$, and $\tau_{49}$ are identically satisfied. This exhaust all the information coming from the coefficients $\tau_{I}$ ($I=0\,,\ldots\,,49$).  Now we only have to determine in an independent way the potential $\Omega(r)$ and the coefficient $C_{2}(r)$, and we have two integration constants: $K_{13}$ and $K_{24}$. The vanishing of the coefficient $C_{2}(r)$ and the constants $K_{13}$ and $K_{24}$ implies the trivial solution: $\Psi^{}_{\rm even}=0\,$.

The only thing left to analyze is the integrability of $C_{2}$, or in other words, the compatibility of the equations for $C'^{}_{2}$ and $C''^{}_{2}$  [Eqs.~\eqref{expdC2-even} and~\eqref{expd2C2-even} respectively].  Another possibility would be to introduce Eq.~\eqref{expression-CE2} into Eq.~\eqref{expd2C2-even}, but in that way we would obtain an integro-differential equation for $\Omega(r)$.  We follow here the first option.  To that end, we compare the derivative of Eq.~\eqref{expdC2-even} with Eq.~\eqref{expd2C2-even}:
\begin{eqnarray}
&&\left( e^{}_2 + e^{}_1\, e^{}_4 - e'^{}_4 - e^2_4 \right) C^{}_2 \nonumber \\
&& + K^{}_{24}\left( e^{}_1\, e^{}_3 + \frac{f}{r^2} - e^{}_3\, e^{}_4 - e'^{}_3 \right) = 0 \,.
\end{eqnarray}
This can be written in a more convenient way as follows
\begin{widetext}
\begin{equation}
\left[\Omega(r)-\Omega^{}_{\ast}(r)\right]\,C^{}_{2}(r) + K^{}_{24}\left\{ e^{}_{5}(r)\,\left[\Omega(r)-\Omega^{}_{\ast}(r)\right]' + e^{}_{6}(r)\,\left[\Omega(r)-\Omega^{}_{\ast}(r)\right] + e^{}_{7}(r) \right\} = 0\,,
\label{expC2-even}
\end{equation}
\end{widetext}
where $\Omega_{\ast}(r)$ is given in Eq.~\eqref{potential-even} and $e^{}_{5}(r)\,$, $e^{}_{6}(r)\,$, and $e^{}_{7}(r)$ are known functions given by:  
\begin{eqnarray}
e^{}_{5}(r) & = & \frac{r\,f^3(r)}{\ell(\ell+1)\lambda(r)} \,, 
\end{eqnarray}
\begin{eqnarray}
e^{}_{6}(r) & = & \frac{2\,r^2f^2(r)}{\ell(\ell+1)}\left(\frac{f(r)}{\lambda(r)r}+\frac{1}{2\,r} \right)' \,, \\
e^{}_{7}(r) & = & \frac{f^2(r)}{r^2} + e^{}_5(r)\left( \Omega^{}_{\ast}(r) - \frac{\lambda'(r)}{r}\right)' \nonumber \\
            & + & e^{}_6(r)\left( \Omega^{}_{\ast}(r) - \frac{\lambda'(r)}{r}\right)\,.
\end{eqnarray}
Equation~\eqref{expC2-even} can in principle provide an expression for the coefficient $C_{2}$ [different from the one obtained in Eq.~\eqref{expression-CE2}]. But it is not guaranteed that such an expression would be compatible with the expressions that we have for its derivatives [Eqs.~\eqref{expdC2-even} and~\eqref{expd2C2-even}]. Given that this equation is the compatibility between $C'^{}_{2}$ and $C''^{}_{2}$, we just need to check the compatibility with $C'^{}_{2}$ [Eq.~\eqref{expdC2-even}]. To that end, we can take the derivative of Eq.~\eqref{expC2-even} and compare with Eq.~\eqref{expdC2-even}.  Actually, if we use Eq.~\eqref{expdC2-even} to eliminate $C'^{}_{2}$ from the derivative of Eq.~\eqref{expC2-even} we obtain a new relation that has the same form as Eq.~\eqref{expC2-even}. In principle, one can repeat this process an arbitrary number of times to get a chain of relations of the form:
\begin{equation}
\Pi^{}_{n}[r,\Omega]\, C^{}_{2}(r) + K^{}_{24}\,\Xi^{}_{n}[r,\Omega] = 0\,,
\end{equation}
where $\Pi^{}_{n}$ and $\Xi^{}_{n}$ are coefficients that depend on the radial coordinate and on the potential function $\Omega(r)$. Using Eq.~\eqref{expdC2-even}, it is easy to find a recurrence for these coefficients:
\begin{eqnarray}
\Pi^{}_{n}[r,\Omega] & = & \Pi'^{}_{n-1}[r,\Omega] + e^{}_{4}(r)\Pi^{}_{n-1}[r,\Omega]\,, 
\label{Pi_n-recurrence}\\
\Xi^{}_{n}[r,\Omega] & = & \Xi'^{}_{n-1}[r,\Omega] + e^{}_{3}(r)\Pi^{}_{n-1}[r,\Omega]\,.
\label{Xi_n-recurrence}
\end{eqnarray}
The case $n=0$ corresponds to Eq.~\eqref{expC2-even} and $n=1$ to its first-order derivative after using Eq.~\eqref{expdC2-even} to eliminate $C'^{}_{2}$. Considering the $n=0$ and $n=1$ equations, there are two possibilities: (i) $C^{}_{2}(r) = K^{}_{24} = 0\,$; (ii) the resultant of the system for $(C^{}_{2}(r),K^{}_{24})$ vanishes, that is:
\begin{equation}
\Pi_{0}\,\Xi_{1}-\Pi_{1}\,\Xi_{0} = 0\,. 
\label{resultant1}
\end{equation}
Given that $\Pi_{0}=\Omega(r)-\Omega^{}_{\ast}(r)$ and that [from Eq.~\eqref{Pi_n-recurrence}] $\Pi_{1}= \Omega'(r)-\Omega'^{}_{\ast}(r)+ e^{}_{4}(r)(\Omega(r)-\Omega^{}_{\ast}(r))$, it is clear that $\Omega(r)=\Omega^{}_{\ast}(r)$ is always a solution. In any case, Eq.~\eqref{resultant1}, becomes an ODE for the potential function $\Omega(r)\,$. In this case, $n=0$, it is a non-linear second-order equation for $\Omega(r)\,$. In principle this equation determines the form of $\Omega(r)$, which in general will be different from $\Omega^{}_{\ast}(r)$ although $\Omega^{}_{\ast}(r)$ is a particular solution. It turns out that the form of this ODE for $\Omega(r)$ is quite similar to the analogous equation that we have obtained in the odd-parity case [see Eq.~\eqref{equation-for-omega-primeprime}].  We find that this ODE for $\delta\Omega(r) = \Omega(r)-\Omega^{}_{\ast}(r)$ is given by
\begin{widetext}
\begin{equation}
K^{}_{24}\left[ \left( \hat{e}^{}_{5}(r) \frac{\delta\Omega'(r)}{\delta\Omega(r)}\right)' + \left(\frac{\hat{e}^{}_{7}(r) }{\delta\Omega(r)}\right)' + \hat{e}'^{}_{6}(r) + \hat{e}^{}_{3}(r) - \delta\Omega(r) \right] = 0 \,,
\label{equation-for-omega-primeprime-even}
\end{equation}
\end{widetext}
where we have introduced several definitions for the coefficients that appear in this equation. First, we have taken advantage that the coefficient $e_4(r)$ can be written as a total derivative [see Eq.~\eqref{coefficient-e4-even}] to introduce the new coefficient $f_4(r)$ as follows:
\begin{eqnarray}
e^{}_4(r) = \left(\ln f^{}_4(r)\right)'\quad\Rightarrow\quad f^{}_4(r) = \frac{r f^2(r)}{\lambda(r)} \,.
\end{eqnarray}
Then, the coefficients that appear in Eq.~\eqref{equation-for-omega-primeprime-even} are defined using $f_4(r)$ in the following way:
\begin{eqnarray}
\hat{e}^{}_{5}(r) & = & \ell(\ell+1)\frac{e^{}_{5}(r)}{f^{}_4(r)} = f(r)\,, 
\label{exp-hat-e5} \\
\hat{e}^{}_{6}(r) & = &  \ell(\ell+1)\frac{e^{}_{6}(r)}{f^{}_4(r)} = 2\lambda(r)r\left(\frac{f(r)}{\lambda(r)r}+\frac{1}{2\,r} \right)' \,, 
\label{exp-hat-e6} \\
\hat{e}^{}_{7}(r) & = &  \ell(\ell+1)\frac{e^{}_{7}(r)}{f^{}_4(r)} =   \ell(\ell+1)\frac{\lambda(r)}{r^3} \nonumber\\
                  & + & \hat{e}^{}_5(r)\left( \Omega^{}_{\ast}(r) - \frac{\lambda'(r)}{r}\right)' \nonumber \\
            & + & \hat{e}^{}_6(r)\left( \Omega^{}_{\ast}(r) - \frac{\lambda'(r)}{r}\right)\,,
\label{exp-hat-e7} \\
\hat{e}^{}_{3}(r) & = &  \ell(\ell+1)\frac{e^{}_{3}(r)|^{}_{\Omega=\Omega_\ast}}{f^{}_4(r)} \nonumber \\
                  & = &  - \left[ \Omega^{}_{\ast}(r) + \frac{\lambda(r) -(\ell+2)(\ell-1)}{r^2} \right]\,.
\label{exp-hat-e3}
\end{eqnarray}
At this point, it is important to remark that the equation for $\delta\Omega(r)$ in the even-parity case, Eq.~\eqref{equation-for-omega-primeprime-even}, has exactly the same structure as the corresponding equation for the odd-parity case [Eq.~\eqref{equation-for-omega-primeprime}]. The only differences are the expressions for the functions of $r$ that appear in them.

If we now consider the next relation in Eqs.~\eqref{Pi_n-recurrence} and~\eqref{Xi_n-recurrence}, namely $n=2$, it can be seen that by combining it with the other two ($n=0$ and $n=1$) we obtain more ODEs for the potential function $\Omega(r)$, this time these ODEs are nonlinear third-order ones. By using the second-order ODE for $\Omega(r)$ that comes from the analysis of the cases $n=0$ and $n=1$, the (two) third-order equations for $\Omega(r)$ are identically satisfied, which ends the analysis. 

As in the odd-parity case, we have to study the two branches that appear. The first branch is characterized by:
\begin{equation}
\Omega(r)=\Omega^{}_{\ast}(r) \,,   
\end{equation}
where $\Omega^{}_{\ast}(r)$ is now given by Eq.~\eqref{potential-even}. Then, Eq.~\eqref{exptau36} is automatically satisfied. Moreover, if we introduce this expression for $\Omega(r)$ into Eq.~\eqref{expC2-even}, we must have that $K^{}_{24}\,e^{}_7(r) = 0\,$. Taking into account that  this has to be valid for any $r$, and that in general $e_7(r)\neq 0$, we must have:
\begin{equation}
K^{}_{24} = 0 \,.
\end{equation}
Then, from Eq.~\eqref{expression-CE2}, we find an expression for the coefficient $C_2$:
\begin{eqnarray}
C^{}_{2}(r) = K^{}_2  \frac{r f^2(r)}{\lambda(r)} \,.
\label{expression-CE2-K24zero}
\end{eqnarray}
This finishes the developments for the first branch. The potential for the even-parity perturbations in this case is given in Eq.~\eqref{potential-even}. It turns out that in the case of a Schwarzschild background, this potential is the well-known Zerilli potential of Eq.~\eqref{schwarzschild-omega-potential-even-parity}. And in the case of a maximally-symmetric background it is the centrifugal barrier potential in Eq.~\eqref{centrifugal-barrier-potential}.
Finally, the most general master function is:
\begin{widetext}
\begin{eqnarray}
\Psi^{}_{\rm even}(t,r) & = & K^{}_2\left\{ \frac{r f^2(r)}{\lambda(r)}h^{}_{11} + \frac{r}{2}\left[\frac{\ell(\ell+1)}{2}G(t,r) + K(t,r)\right] -\frac{f(r)}{\lambda(r)}\left[ \ell(\ell+1)\jeven^{}_1(t,r) + r^2 K'(t,r)\right] \right\} \nonumber \\
& + & \frac{K^{}_{13}}{\lambda(r)}\left\{ r\left( \dot{K}(t,r) + \frac{\lambda(r)}{2}\dot{G}(t,r) - \frac{f'(r)}{r}\jeven^{}_0\right) + f(r)\left(h^{}_{01}(t,r) - \jeven'^{}_0(t,r) + \dot{\jeven}^{}_1(t,r)\right) \right\} \,. 
\end{eqnarray}
\end{widetext}
Using the definitions introduced in Sec.~\ref{master-functions} we can rewrite the master function as follows:
\begin{eqnarray}
\Psi^{}_{\rm even}(t,r) =  \frac{\ell(\ell+1)}{4}K^{}_2\,\Psi^{}_{\rm ZM}(t,r)  - K^{}_{13}\,\Psi^{}_{\rm EN}(t,r) \,,
\label{master-function-even-second-branch}
\end{eqnarray}
where $\Psi^{}_{\rm ZM}(t,r)$ is the Zerilli-Moncrief master function given in Eq.~\eqref{zerilli-moncrief-master-function} and $\Psi^{}_{\rm EN}(t,r)$ is another master function (not known as far as we can say) given by:
\begin{widetext}
\begin{equation}
\Psi^{}_{\rm EN}(t,r) =   \frac{r}{\lambda(r)}\left\{  
\frac{f(r)}{r}\left[ \jeven'^{}_0(t,r) - h^{}_{01}(t,r) -  \dot{\jeven}^{}_1(t,r)  \right] 
+ \frac{f'(r)}{r}\jeven^{}_0  - \dot{K}(t,r) - \frac{\lambda(r)}{2}\dot{G}(t,r)  \right\}     
\label{new-even-function}
\end{equation}
\end{widetext}
which can be rewritten in a completely covariant form as follows:
\begin{equation}
\Psi^{}_{\rm EN}(t,r) = \frac{1}{\lambda(r)}t^a\left(r \tilde{K}^{}_{:a} - \tilde{h}^{}_{ab}r^b \right) \,,  
\label{new-even-parity-master-function}
\end{equation}
where $\tilde{h}^{}_{ab}$ and $\tilde{K}$ are the gauge-invariant quantities introduced in Eqs.~\eqref{expression-hathab} and~\eqref{expression-hatK} respectively.  Hence, $\Psi^{}_{\rm EN}(t,r) $ is also a gauge-invariant master function that can be written in a covariant way. By using the perturbative field equations it is possible to show that the master function $\Psi^{}_{\rm EN}(t,r)$ turns out to be proportional to the time derivative of the Zerilli-Moncrief master function: 
\begin{equation}
t^a \Psi^{}_{{\rm ZM} :a} = 2\,\Psi^{}_{\rm EN}\,. 
\label{ZMdot-EN}
\end{equation}
This ends the analysis of the first branch.  
 
In the second branch, when $\Omega(r)\neq\Omega^{}_{\ast}(r)$, by virtue of Eq.~\eqref{exptau36} we must have
\begin{equation}
K^{}_{13}= 0\,.    
\end{equation}
In this case, Eq.~\eqref{expC2-even} provides a second expression for $C_2(r)$ that has to be compared with the other expression, Eq.~\eqref{expression-CE2}. The outcome of this comparison is the second-order non-linear ODE for $\Omega(r)$ that is given in Eq.~\eqref{equation-for-omega-primeprime-even}. Moreover, in this branch, the only non-zero coefficients of $\Psi^{}_{\rm even}$ are: $C_{2}(r)$ [Eq.~\eqref{expression-CE2} with $\Omega(r)$ satisfying the non-linear ODE of Eq.~\eqref{equation-for-omega-primeprime-even}], $C_{4}(r)$ [Eq.~\eqref{expression-for-C4-even}], $C_{5}(r)$ [Eq.~\eqref{expression-for-C5-even}], $C_{6}(r)$ [Eq.~\eqref{expression-for-C6-even}], $C_{18}(r)$ [Eq.~\eqref{expression-for-C18-even}], and $C_{20}$ [Eq.~\eqref{expression-for-C20-even}]. Introducing these expressions into $\Psi^{}_{\rm even}$ we can write the result in the following form:
\begin{widetext}
\begin{equation}
\Psi^{}_{\rm even}(t,r) = \left( K^{}_2 + {K}^{}_{24}\,\Sigma(r) \right) \Psi^{}_{\rm ZM}(t,r)  - \frac{K^{}_{24}}{2}\,\tilde{K}(t,r) \,,
\end{equation}
\end{widetext}
where:
\begin{eqnarray}
\Sigma = \frac{1}{\ell(\ell+1)} \left[ \frac{\lambda - (\ell+2)(\ell-1)}{2\,r} - \int^{}_r dr'\Omega(r') \right] \,,
\end{eqnarray}
and $\tilde{K}(t,r)$ is the gauge-invariant combination of metric perturbations introduced in Eq.~\eqref{expression-hatK}. Therefore, the even-parity master function in this second branch is also gauge invariant. Actually, it is a linear combination of the Zerilli master function and another master function that is a combination of the Zerilli master function and $\tilde{K}(t,r)$, which depends on the potential $\Omega(r)$. The potential, in turn, satisfies the non-linear ODE in Eq.~\eqref{equation-for-omega-primeprime-even}. As in the odd-parity case, this equation can be simplified [see Eq.~\eqref{xdarboux-odd}]. To that end, we  have to write it in terms of the potential difference $\delta V = f\delta\Omega$ [see Eq.~\eqref{potential-rr*}], use the expressions for $\hat{e}_3$ and $\hat{e}_6$, which satisfy
\begin{equation}
\hat{e}^{}_3 + \hat{e}'^{}_6 - f'' = 0\,,
\end{equation}
and finally, we have to exchange derivatives with respect to $r$ with derivatives with respect to the tortoise coordinate. After doing all this, we arrive at a simplified equation that looks exactly like the one for the odd-parity case [see Eq.~\eqref{xdarboux-odd}]
\begin{equation}
\left(\frac{\delta{V}^{}_{,x}}{\delta{V}} \right)^{}_{,x}
+ 2 \left(\frac{V^{\rm even}_{,x}}{\delta{V}} \right)^{}_{,x} - \delta{V} = 0 \,,
\label{xdarboux-even}
\end{equation}
where $V^{\rm even}_{} = f\Omega_{\ast}$ is the Zerilli potential.  This ends the analysis of the even-parity case.

\section{Conclusions and Future Prospects}\label{Conclusions-Future-Prospects}

In this paper, we have carried out a study of all the possible master functions and equations for the perturbations of vacuum spherically-symmetric spacetimes. The only assumptions made in this study are: (i) The master functions are linear combinations of the metric perturbations and their first-order derivatives. (ii) The coefficients of those linear combinations only depend on the radial areal coordinate. (iii) The master functions satisfy a wave-type equation associated with the two-dimensional metric of the Lorentzian manifold tangent to the spheres of symmetry, and with a potential that is determined by the perturbative Einstein equations. 

The outcome of this study produces two branches of solutions: (a) The first branch corresponds to the already known results, with the exception of the even-parity case, for which we have found a new master function independent of the Zerilli-Moncrief one, $\Psi_{\rm ZM}$, and which we have denoted by $\Psi_{\rm EN}$. For both parities, the most general master function is a linear combination (with constant coefficients) of two independent master functions. These master functions can be taken to be the Regge-Wheeler and the Cunningham-Prince-Moncrief master functions, $(\Psi_{\rm RW},\Psi_{\rm CPM})$, in the odd-parity case and $(\Psi_{\rm ZM},\Psi_{\rm EN})$ in the even-parity case.  One the other hand, the potentials are the known ones: The Regge-Wheeler (odd-parity) and the Zerilli (even-parity) potentials.
(b) The second branch was essentially unknown and, in contrast with the first branch, there are infinite possible potentials, different from the ones already known (first branch). The set of possible potentials corresponds to the solutions of a non-linear differential equation which has the same form for both parities [see Eqs.~\eqref{xdarboux-odd} and~\eqref{xdarboux-even}]. The master functions are again a linear combination (with coefficients depending only on the radial area coordinate) of two independent mas ter functions. In the odd-parity case, they can be taken to be the Cunningham-Price-Moncrief master function and a new one that is a combination of $\Psi_{\rm CPM}$ and $\Phi_{\rm ON}$ that includes the potential function [see Eqs.~\eqref{master-function-odd-second-branch} and~\eqref{odd-new-function}].  The even-parity case follows the same pattern, and the most general master function can be taken to be a linear combination of the Zerilli-Moncrief master function and another new master function made out of a combination of $\Psi_{\rm ZM}$, $\Phi_{\rm EN}$, and the potential function [see Eqs.~\eqref{master-function-even-second-branch} and~\eqref{new-even-parity-master-function}].

Apart from the construction of the master functions and equations (potentials), it is important to remark other findings that came out from our developments: 
(i) The flow of the argument is the same for the two parities despite the different number of variables and equations that describe the metric perturbations in the two cases. 
(ii) Gauge invariance: All the master functions and equations turn out to be gauge-invariant, which is something that we did not impose. In this sense, it is important to remark that we have always worked on a general gauge. The emerging gauge invariance is then due to the physical/geometric character of the (master) wave equations. This shows the important role played by the master functions, which in some sense encode the true degrees of freedom of the (perturbative) gravitational field. 
(iii) Despite the fact that in many places we have carried out the calculations using a  specific class of coordinate systems, we have been able always to restore full covariance with respect to the 1+1 Lorentzian metric $g_{ab}$ [see Eq.~\eqref{warped-metric-background}]. 
(iv) For both parities, in the first branch, one of the independent master functions can be taken to be the time derivative of the other one [see Eqs.~\eqref{CPMdot-RW} and~\eqref{ZMdot-EN}]. 
(v) In the case when the background is maximally symmetric, the odd-parity and even-parity potentials of the first branch are identical. In the case of the second branch, the set of possible potentials are the same for the two parities.  This means that the maximal symmetry has a strong impact on the possible set of master equations and functions.  There are still two branches, but they are identical for both parities. 
(vi) We have always worked in the time domain. This has the advantage that at any moment we can obtain results in the frequency domain by introducing the standard substitution: $\Psi(t,r)\rightarrow e^{i\omega t}\phi(r)$.

Our analysis is quite general, in the sense that it is based on a few assumptions (see Sec.~\ref{construction-master-functions}), and complete, in the sense that it unveils the full content of the (first-order) perturbative approach to vacuum spherically symmetric spacetimes. Within our knowledge, no similar analysis has been carried out before. The systematic construction of master functions and equations we have followed can be applied directly to other different scenarios in spherical symmetry, in particular to systems involving matter fields (see~\cite{Gerlach:1979rw,Gerlach:1980tx} for a general approach): point particle~\cite{Davis:1971gg}, electromagnetic fields (see, e.g.~\cite{Moncrief:1974gw,Moncrief:1974ng,Moncrief:1975sb,Chandrasekhar:1979iz,Chandrasekhar:10.2307/79860,Xanthopoulos:10.2307/2397164}), perfect fluids~\cite{Gundlach:1999bt,Martin-Garcia:2000ze},  etc.  Within General Relativity it can also be applied to spacetimes with a different number of dimensions, in particular it would be interesting to study the case of three spacetime dimensions, where we have the Ba\~nados-Teitelboim-Zanelli (BTZ) Black Hole~\cite{Banados:1992wn} (which is asymptotically AdS${}_{3}$), and for which there are analytic solutions for the quasinormal frequencies and wave functions~\cite{Ichinose:1994rg}. Again within General Relativity, one can try to follow the procedure for second- and higher-order perturbations~\cite{Gleiser:1995gx,Gleiser:1998rw,Brizuela:2006ne} of spherically-symmetric spacetimes. This is particularly interesting taking into account that the equations for higher-order perturbations usually contain the same differential operators of the background as the ones for the first-order perturbations. Finally, this procedure can also be applied to perturbations of spherically-symmetric spacetimes in other theories of gravity (see, e.g.~\cite{Kobayashi:2012kh,Kobayashi:2014wsa}).

Another interesting question is whether we can apply a similar procedure in the case of the Kerr metric~\cite{Teukolsky:1973ha} and other axially-symmetric spacetimes. That would make contact with the Bardeen-Press master function~\cite{Bardeen:1973xb} for the Schwarzschild spacetime, which is not included in our analysis since it contains second-order derivatives of the metric perturbations. In any case, it is clear that for Kerr perturbations we should allow for the presence of second-order derivatives.

On the other hand, the results of our study lead to a number of questions. In particular, what is the meaning of the infinite set of possible master functions and equations that appear in the second branch of solutions.  We clarify this question in a forthcoming paper~\cite{Lenzi:2021njy}, where we analyze in detail the connection between this plethora of master equations and master functions and show that all of them are related by Darboux transformations. Nevertheless, the Darboux transformation that connects them has to be interpreted in a more general context than the classical Darboux transformation which is normally introduced in the context of Sturm-Liouville problems, with self-adjoint operators. In this sense, in~\cite{Lenzi:2021njy} we show the crucial role played by the equations for the potentials of the second branch, equations~\eqref{xdarboux-odd} and~\eqref{xdarboux-even}.

\begin{acknowledgments}
We thank Jos\'e Lu\'{\i}s Jaramillo for enlightening discussions regarding the dynamics of perturbed black holes.
CFS is supported by contracts ESP2017-90084-P and PID2019-106515GB-I00/AEI/10.13039/501100011033 (Spanish Ministry of Science and Innovation) and 2017-SGR-1469 (AGAUR, Generalitat de Catalunya). 
We thank the COST Action CA16104 Gravitational waves, black holes and fundamental physics (GWverse) as a forum of discussion where part of this work was originated and for a Short Term Scientific Mission award to ML that allowed us to complete this work.
ML also acknowledges support from an “Angelo della Riccia” fellowship. 
We have used the Mathematica software~\cite{Mathematica} to implement the mathematical machinery related with Einstein's equations for spherically-symmetric spacetimes shown in the Appendices. Using this Mathematica implementation we have checked most of the computations necessary for this paper. 
\end{acknowledgments}

\appendix

\section{Differential Properties of Spherical Harmonics}\label{sphericalharmonics}
All the spherical harmonics (scalar, vector, and tensor) we use in this paper are determined once we prescribe the scalar spherical harmonics (see, e.g.~\cite{Abramowitz:1970as,Press:1992nr}). The vector and tensor spherical harmonics used in this paper can then be obtained via equations~(\ref{vectorharmonics})-(\ref{tensorharmonics}). To recover the equations and results of this paper, we only need to use some differential identities that they satisfy:  The even-parity vector harmonics $Y^{\ell m}_{A}$ satisfy the following differential identities:
\begin{eqnarray}
\Omega^{AB} Y^{\ell m}_{A|B} & = & -\ell(\ell+1)Y^{\ell m} \,, \\
\Omega^{BC} Y^{\ell m}_{B|CA} & = & -\ell(\ell+1)Y^{\ell m}_{A} \,,\\
\Omega^{BC} Y^{\ell m}_{A|BC} & = & \left[1 -\ell(\ell+1)\right]Y^{\ell m}_{A} \,.
\end{eqnarray}
The odd-parity vector harmonics $X^{\ell m}_{A}$ satisfy similar differential identities:
\begin{eqnarray}
\Omega^{AB} X^{\ell m}_{A|B} & = & 0 \,, \\
\Omega^{BC} X^{\ell m}_{A|BC} & = & \left[1 -\ell(\ell+1)\right]X^{\ell m}_{A} \,,\\
\Omega^{BC} X^{\ell m}_{B|AC} & = & X^{\ell m}_{A} \,.
\end{eqnarray}
On the other hand, the (symmetric) even-parity tensor harmonics $T^{\ell m}_{AB}$ and $Y^{\ell m}_{AB}$ satisfy the following differential identities:
\begin{eqnarray}
\Omega^{BC} T^{\ell m}_{BC|A} & = & 2\,Y^{\ell m}_{A} \,, \\
\Omega^{BC} T^{\ell m}_{AB|C} & = & Y^{\ell m}_{A} \,,\\
\Omega^{CD} T^{\ell m}_{AB|CD} & = & -\ell(\ell+1) T^{\ell m}_{AB} \,,\\
\Omega^{CD} T^{\ell m}_{CD|AB} & = & Y^{\ell m}_{AB} -\ell(\ell+1) T^{\ell m}_{AB} \,.
\end{eqnarray}
\begin{eqnarray}
\Omega^{BC} Y^{\ell m}_{BC|A} & = & 0 \,, \\
\Omega^{BC} Y^{\ell m}_{AB|C} & = & -\frac{(\ell+2)(\ell-1)}{2} Y^{\ell m}_{A} \,,\\
\Omega^{CD} Y^{\ell m}_{AB|CD} & = & \left[4 -\ell(\ell+1)\right] Y^{\ell m}_{AB} \,.
\end{eqnarray}
Finally, the (symmetric) odd-parity tensor harmonics $X^{\ell m}_{AB}$ satisfy the following differential identities:
\begin{eqnarray}
\Omega^{BC} X^{\ell m}_{BC|A} & = & 0 \,, \\
\Omega^{BC} X^{\ell m}_{AB|C} & = & -\frac{(\ell+2)(\ell-1)}{2} X^{\ell m}_{A} \,,\\
\Omega^{CD} X^{\ell m}_{AB|CD} & = & \left[4 -\ell(\ell+1)\right] X^{\ell m}_{AB} \,.
\end{eqnarray}
%

\section{Multipolar Components of Geometric Perturbative Quantities}\label{multipolar-components-perturbations}
We give expressions of the main quantities that we need to analyze the perturbative vacuum (with cosmological constant) Einstein equations (see~\cite{Martel:2005ir} for complementary expressions). 
The components of the perturbation of the Christoffel symbols (these are tensors from the point of view of the background spacetime), introducing the spherical harmonic decomposition of the metric perturbations into Eq.~\eqref{perturbed-Christoffel} are:
\begin{widetext}
\begin{eqnarray}
\delta\Gamma^a{}^{}_{bc} & = & \frac{1}{2} g^{ad}\left( h^{}_{cd:b} - h^{}_{bc:d} + h^{}_{bd:c}\right)Y \,, 
\label{dGabc} \\
\delta\Gamma^a{}^{}_{bA} & = & \left( \frac{1}{2}p^a{}^{}_b  + g^{ac}\jeven^{}_{[c:b]} - \frac{r^{}_b}{r}\jeven^a \right)Y^{}_A + 
                               \left( g^{ac}h^{}_{[c:b]} - \frac{r^{}_b}{r}h^a\right)X^{}_A\,, 
\label{dGabA} \\
\delta\Gamma^a{}^{}_{AB} & = & \left(r r^b\, p^a{}^{}_b - \frac{\ell(\ell+1)}{2}\jeven^a - \frac{1}{2}\left(r^2 K\right)^{:a}\right)T^{}_{AB}
                              +\left(\jeven^a - \frac{1}{2}\left(r^2 G\right)^{:a}\right)Y^{}_{AB}
                              +\left(h^a - \frac{1}{2}h^{}_2{}^{:a}\right)X^{}_{AB} \,, 
\label{dGaAB} \\
\delta\Gamma^A{}^{}_{ab} & = & \frac{1}{r^2}\left( \jeven^{}_{a:b}-\frac{1}{2}p^{}_{ab}\right)Y^A 
                              +\frac{1}{r^2}h^{}_{a:b} X^A \,, 
\label{dGAab} \\
\delta\Gamma^A{}^{}_{Ba} & = & \frac{1}{2 r^2}\left( \left(r^2 K\right)^{}_{:a} - 2\, r r^{}_a K \right) T^A{}^{}_B 
                              +\frac{1}{2 r^2}\left( \left(r^2 G\right)^{}_{:a} - 2\, r r^{}_a G \right) Y^A{}^{}_B
                              +\frac{1}{2 r^2}\left( h^{}_{2:a} -2 \frac{r^{}_a}{r}h^{}_2 \right)X^A{}^{}_B \nonumber \\
                         & + & \frac{\ell(\ell+1)}{2 r^2}h^{}_a \, Y \epsilon^A{}^{}_B \,.  
\label{dGABa} \\
\delta\Gamma^A{}^{}_{BC} & = & \frac{1}{2} K \left( \Omega^A{}^{}_B Y^{}_C + \Omega^A{}^{}_C Y_B -\Omega^{}_{BC}Y^A \right)
                              +\frac{1}{2} G \,\Omega^{AD}\left( Y^{}_{CD|B} - Y^{}_{BC|D} + Y^{}_{BD|C} \right) \nonumber \\
                         & + & \frac{h^{}_2}{2 r^2} \Omega^{AD}\left( X^{}_{CD|B} - X^{}_{BC|D} + X^{}_{BD|C} \right)
                              +\frac{r^a}{r} \left( \jeven^{}_a\, Y^A + h^{}_{a}\, X^A \right) \Omega^{}_{BC} \,.
\label{dGABC}
\end{eqnarray}
\end{widetext}
Now, in the same way that we have the harmonic decomposition of the perturbed Christoffel symbols in terms of the metric perturbation harmonics, we can write the harmonic decomposition of the Riemann tensor in terms of the harmonic decomposition of the perturbations of the Christoffel symbols.  Then, from Eq.~\eqref{perturbed-Riemann} we obtain:
\begin{widetext}
\begin{eqnarray}
\delta R^a{}^{}_{bcd} & = & \delta\Gamma^a{}^{}_{bd:c} - \delta\Gamma^a{}^{}_{bc:d} \,, \\
\delta R^a{}^{}_{bcA} & = & \delta\Gamma^a{}^{}_{bA:c} - \delta\Gamma^a{}^{}_{bc|A} + r r^a \Omega^{}_{AB} \delta\Gamma^B{}^{}_{bc} 
                           +\frac{r^{}_b}{r} \delta\Gamma^a{}^{}_{Ac}\,, \\
\delta R^a{}^{}_{bAB} & = & \delta\Gamma^a{}^{}_{bB|A} - \delta\Gamma^a{}^{}_{bB|A} - r r^a\left( \Omega^{}_{AC}\delta\Gamma^C{}^{}_{bB}
                           -\Omega^{}_{BC}\delta\Gamma^C{}^{}_{bA} \right) \,, \\
\delta R^a{}^{}_{Abc} & = & \delta\Gamma^a{}^{}_{cA:b} - \delta\Gamma^a{}^{}_{bA:c} - \frac{r^{}_b}{r} \delta\Gamma^a{}^{}_{cA}
                           +\frac{r^{}_c}{r}\delta\Gamma^a{}^{}_{bA}\,, \\
\delta R^a{}^{}_{AbB} & = & \delta\Gamma^a{}^{}_{AB:b} - \delta\Gamma^a{}^{}_{bA|B} - \frac{r^{}_b}{r} \delta\Gamma^a{}^{}_{AB}
                           + r r^a \Omega^{}_{BC}\delta\Gamma^C{}^{}_{Ab} - r  r^c \delta \Gamma^a{}^{}_{bc}\Omega^{}_{AB}\,, \\
\delta R^a{}^{}_{ABC} & = & \delta\Gamma^a{}^{}_{AC|B} - \delta\Gamma^a{}^{}_{AB|C} 
                           - r r^a \left( \Omega^{}_{BD}\delta\Gamma^D{}^{}_{AC} - \Omega^{}_{CD}\delta\Gamma^D{}^{}_{AB}\right) 
                           + r r^b \left( \Omega^{}_{AB}\delta\Gamma^a{}^{}_{bC} - \Omega^{}_{AC}\delta\Gamma^a{}^{}_{bB}\right)  \,,  \\  
\delta R^A{}^{}_{abc} & = & \delta\Gamma^A{}^{}_{ac:b} - \delta\Gamma^A{}^{}_{ab:c} + \frac{r^{}_b}{r} \delta\Gamma^A{}^{}_{ac}
                           -\frac{r^{}_c}{r}\delta\Gamma^A{}^{}_{ab}\,, \\
\delta R^A{}^{}_{abB} & = & \delta\Gamma^A{}^{}_{aB:b} - \delta\Gamma^A{}^{}_{ab|B} + \frac{r^{}_a}{r} \delta\Gamma^A{}^{}_{Bb}
                           +\frac{r^{}_b}{r}\delta\Gamma^A{}^{}_{Ba} - \frac{r^{}_c}{r}\delta\Gamma^c{}^{}_{ab} \delta^A{}^{}_B \,, \\
\delta R^A{}^{}_{aBC} & = & \delta\Gamma^A{}^{}_{aC|B} - \delta\Gamma^A{}^{}_{aB|C} 
                           +\frac{r^{}_b}{r}\left( \delta\Gamma^b{}^{}_{aC}\delta^A{}^{}_B - \delta\Gamma^b{}^{}_{aB}\delta^A{}^{}_C \right)\,, \\
\delta R^A{}^{}_{Bab} & = & \delta\Gamma^A{}^{}_{Bb:a} - \delta\Gamma^A{}^{}_{Ba:b} \,, \\
\delta R^A{}^{}_{BaC} & = & \delta\Gamma^A{}^{}_{BC:a} - \delta\Gamma^A{}^{}_{Ba|C} - \frac{r^{}_b}{r}\delta\Gamma^b{}^{}_{aB}\delta^A{}^{}_C
                           - r r^{b} \delta\Gamma^A{}^{}_{ab}\Omega^{}_{BC}\,, \\
\delta R^A{}^{}_{BCD} & = & \delta\Gamma^A{}^{}_{BD|C} - \delta\Gamma^A{}^{}_{BC|D} 
                           +\frac{r^{}_a}{r}\left( \delta\Gamma^a{}^{}_{BD}\delta^A{}^{}_C - \delta\Gamma^a{}^{}_{BC}\delta^A{}^{}_D \right)
                           + r r^a\left( \Omega^{}_{BC}\delta\Gamma^A{}^{}_{aD} - \Omega^{}_{BD}\delta\Gamma^A{}^{}_{aC} \right) \,.
\end{eqnarray}
\end{widetext}
From here we can find the harmonic decomposition of the perturbations of the Ricci tensor by introducing the perturbations of the Christoffel symbols in terms of the metric perturbation harmonics [Eqs~\eqref{dGabc}-\eqref{dGABC}]. The result is
\begin{widetext}
\begin{eqnarray}
\delta R^{}_{ab} & = & \left[ h^{}_{c (a:b)}{}^{:c} - \frac{1}{2}h^{}_{ab:c}{}^{:c} - \frac{1}{2}h^{}_{:ab}
                            + \frac{r^c}{r} \left( h^{}_{ac:b} + h^{}_{bc:a} - h^{}_{ab:c} \right) 
                            - K^{}_{:ab} - \frac{2}{r}r^{}_{(a}K^{}_{:b)}
                            + \frac{\ell(\ell+1)}{2 r^2} \left( h^{}_{ab} - 2\,\jeven^{}_{a:b}\right)\right]Y \,, \\
\delta R^{}_{aA} & = & \left[ \frac{1}{2}h^{}_{ab}{}^{:b} - \frac{1}{2}h^{}_{:a} + \frac{r^{}_a}{2 r}h + \frac{1}{2}\jeven^c{}^{}_{:ca}
                            - \frac{1}{2}\jeven^{}_{a:c}{}^c - \frac{r^{}_a}{r}\jeven^b{}^{}_{:b} + \frac{r^b}{r}\jeven^{}_{b:a}
                            + \left(\frac{{}^{2}R}{4} g^{}_{ab} - \frac{r^{}_{:ab}}{r}-\frac{r^{}_a r^{}_b}{r^2}\right)\jeven^b
                            - \frac{1}{2}K^{}_{:a} \right. \nonumber \\
                 &   & \left. - \frac{(\ell+2)(\ell-1)}{4} G^{}_{:a}\right]Y^{}_A
                     + \left[ \frac{1}{2}h^c{}^{}_{:ca} - \frac{1}{2}h^{}_{a:c}{}^c - \frac{r_a}{r} h^b{}^{}_{:b}
                            + \frac{r^b}{r}h^{}_{b:a} + \left(\frac{{}^{2}R}{4} g^{}_{ab} - \frac{r^{}_{:ab}}{r}-\frac{r^{}_a r^{}_b}{r^2}\right) h^b
                            + \frac{\ell(\ell+1)}{2 r^2} h^{}_a \right. \nonumber \\
                 &   & \left. - \frac{(\ell+2)(\ell-1)}{4 r^2}\left(h^{}_{2:a}-\frac{2 r^{}_a}{r}h^{}_2 \right)\right]X^{}_A\,, \\
\delta R^{}_{AB} & = & \left[ r r^c h^a{}^{}_{c:a} - \frac{r}{2}r^a h^{}_{:a} +\left(r^a r^b + r r^{:ab}\right) h^{}_{ab}
                            + \frac{\ell(\ell+1)}{4} h - \frac{\ell(\ell+1)}{2} \left( \jeven^a{}^{}_{:a} + 2\,\frac{r^a}{r}\jeven^{}_a\right)
                            - \frac{1}{2} \left(r^2 K\right)^{:a}{}^{}_{:a} \right. \nonumber \\
                 &   & \left. + \frac{\ell(\ell+1)}{2} K + \frac{(\ell+2)(\ell+1)\ell(\ell-1)}{4}G \right] T^{}_{AB}
                      +\left[ \jeven^a{}^{}_{:a} - \frac{1}{2}r^2 G^{}_{:a}{}^{:a} - r r^a G^{}_{:a} - \left( r r^{:a}{}^{}_{:a} + r^a r^{}_a - 1 \right) G
                            - \frac{h}{2}\right] Y^{}_{AB} \nonumber \\
                 & + & \left[ h^a{}^{}_{:a} - \frac{1}{2}h^{}_{2:a}{}^{:a} + \frac{r^a}{r} h^{}_{2:a} + \frac{\left(1-2 r^a r^{}_a\right)}{r^2} h^{}_2 \right] X^{}_{AB}\,. 
\end{eqnarray}
\end{widetext}
%


%

\end{document}